\RequirePackage{fix-cm}
\documentclass[twocolumn]{svjour3}          
\smartqed  
\usepackage{booktabs} 
\usepackage{amsfonts,amssymb}
\usepackage[linesnumbered,ruled,vlined]{algorithm2e}
\usepackage{amsmath}
\usepackage{graphicx}
\usepackage{subcaption}
\usepackage{float}
\usepackage{threeparttable}
 \usepackage{hyperref} 
\usepackage{comment}
\usepackage{adjustbox}   
\usepackage{graphicx}
\usepackage{enumitem}
 \usepackage{multirow}
\usepackage{comment}
\usepackage{pdflscape}
\usepackage{cite}   
\usepackage{xcolor} 

\newcommand{\methodG}{$\text{GeSI}$\xspace}

\newcommand{\methodE}{$\text{EmSI}$\xspace}

\begin{document}

\title{Conceptual Schema Inference for Tabular Datasets using Large Language Models
}


\author{Zhenyu Wu         \and
        Jiaoyan Chen \and 
        Norman W. Paton
}


\institute{Zhenyu Wu, Jiaoyan Chen, Norman W. Paton \at
              The University of Manchester, Manchester, UK\\
              \email{\{zhenyu.wu,jiaoyan.chen,norman.paton\}@manchester\\.ac.uk}           
}

\date{Received: date / Accepted: date}

\maketitle

\begin{abstract}
Large collections of tabular data from data lakes, web tables and open data portals often originate from heterogeneous sources, leading to representational inconsistencies. Understanding and organizing such repositories therefore remains a major challenge. While prior work has primarily focused on dataset discovery and exploration, this paper addresses the complementary problem of conceptual schema inference: automatically deriving a conceptual schema that captures entity types, attributes and inter-type relationships directly from raw tables. 
We propose two large language model (LLM)-based approaches that use only column headers and cell values: \methodG uses generative LLMs to infer hierarchical types and their attributes from table- and column-level semantics, and to integrate them into a global schema that also captures relationships across types; \methodE employs LLM-based table embeddings to group tables by column-level semantics, infer attributes within each group, and construct hierarchical structures from shared attribute patterns.
Finally, we report an experimental analysis demonstrating the effectiveness of our approaches in terms of the conciseness and structural quality of the inferred schema components, their scalability to large repositories, and a case study illustrating end-to-end schema inference.

\keywords{Conceptual Schema Inference\and Large Language Models \and Prompt Learning \and Table Embedding }
\end{abstract}

\section{Introduction}
\label{intro} 

Data lakes, web tables and open data provide users with large repositories of minimally curated datasets with the potential for analysis~\cite{DBLP:journals/tkde/HaiKQJ23,DBLP:journals/tist/ZhangB20,DBLP:conf/chi/LiuUZS23}. Such repositories typically bring together data from different producers or publishers, and consequently manifest schema and instance level heterogeneities at scale. Because of these features, it may be difficult for data scientists to obtain an understanding of what data is available that could potentially contribute to an analysis at hand.

The difficulty in obtaining value from complex data repositories has been recognized for some time, and a variety of techniques have been developed to support data discovery and exploration~\cite{DBLP:journals/csur/PatonCW24}. For example, dataset search can be based on keywords~\cite{conf/VLDB/CafarellaOctopus09,conf/WWW/DanBrickey19} or {can use} datasets as queries~\cite{DBLP:journals/pvldb/NargesianZPM18,DBLP:conf/icde/BogatuFP020}; given a dataset, users can navigate to related datasets by exploring various kinds of relationships~\cite{DBLP:conf/icde/FernandezMQEIMO18,ACM:proceedings:JOSIESigmod19}; and when suitable ontologies are available, annotations can be inferred for datasets or their attributes~\cite{DBLP:conf/ijcai/ChenJHS19,DBLP:journals/pvldb/DengSL0020}. While effective for retrieving relevant datasets, these methods typically operate at the level of individual tables or attributes, and thus fall short of providing users with a global, conceptual understanding of an entire repository. For example, which kinds of types the datasets are about, what attributes commonly appear for those types, and what plausible relationships recur across them. 

\begin{figure*}[tb]
  \centering

  \begin{subfigure}[t]{0.485\textwidth}
 \centering
      \resizebox{\textwidth}{!}{ 
    \begin{tabular}{|l|l|l|l|l|} \hline
\textbf{Event\_title} & \textbf{composer} & \textbf{performer} & \textbf{venue} & \textbf{event\_date} \\
\hline
Carmen & Bizet & Elīna Garanča & The Metropolitan Opera & 2022-10-01 \\\hline
Mozart Requiem & Mozart & Salzburg Festival Choir & Großes Festspielhaus & 2023-08-12 \\\hline
Tosca	 &Puccini &	Anna Netrebko &	Teatro alla Scala	 &2023-11-05\\\hline
Beethoven Symphony No.5	&Beethoven	 &Berlin Philharmonic	 &Philharmonie Berlin &	2024-03-22\\
 No.5	Beethoven	 &	 & &	&\\\hline
\end{tabular}}
    \caption{d1: Music Event}
    \label{tab:musicEvent}

  \end{subfigure}%
  \begin{subfigure}[t]{0.50\textwidth}
    \centering
  \resizebox{\textwidth}{!}{
\begin{tabular}{|l|l|l|l|l|} \hline
\textbf{album\_title} & \textbf{composer} & \textbf{performer} & \textbf{venue} & \textbf{release\_date} \\ \hline
 Vienna Gala Concert & Strauss & Vienna Philharmonic & Musikverein Vienna & 2024-03-15 \\ \hline
Mozart Requiem & Mozart & Salzburg Festival Choir & Großes Festspielhaus & 2023-10-17 \\ \hline
Carmen & Bizet & Elīna Garanča & The Metropolitan Opera & 2022-11-25 \\ \hline
Les Misérables & Schönberg & London Cast Ensemble & Queen’s Theatre London & 2023-09-08 \\ \hline
\end{tabular}
}
   \caption{d2: Music Recording}
    \label{tab:musicRecording}
  \end{subfigure}

  \begin{subfigure}[t]{0.49\textwidth}
    \centering
    \resizebox{0.8\textwidth}{!}{ 
\begin{tabular}{|l|l|l|l|l|}
\hline
\textbf{name} & \textbf{location} & \textbf{established\_year} & \textbf{rating} & \textbf{source} \\ \hline
Opera House Hotel & New York & 2013 & 4.5 & Booking \\ \hline
The Queen's Gate Hotel & London & 1857 & 4.3 & Expedia \\ \hline
Hotel am Opernring & Vienna & N/A & 4.4 & Airbnb \\ \hline
Hotel Opera & Munich & 1968 & 4.4 & Tripadvisor \\ \hline
\end{tabular}
}
  \caption{d3: Hotel}
   \label{tab:Hotel}

  \end{subfigure}%
  \hfill
 \begin{subfigure}[t]{0.49\textwidth}
    \centering
          \resizebox{\textwidth}{!}{ 
\begin{tabular}{|l|l|l|l|} \hline
\textbf{name} & \textbf{location} & \textbf{established\_year} & \textbf{rating} \\ \hline
The Metropolitan Opera & Lincoln Center District, New York, USA & 1883 & 4.8 \\ \hline
Queen’s Theatre London & London, UK & 1907 & 4.7 \\ \hline
Vienna State Opera & Vienna, Austria & 1869 & 4.9 \\ \hline
Bavarian State Opera & Munich, Germany & 1653 & 4.8 \\ \hline
\end{tabular}}
\caption{d4: Opera House}
   \label{tab:OperaHouse}
  \end{subfigure} 

  \begin{subfigure}[t]{0.47\textwidth}
    \centering
  \resizebox{\textwidth}{!}{ 
  \begin{tabular}{|l|l|l|l|l|}
  \hline  \textbf{date} & \textbf{start\_city} & \textbf{organizer} & \textbf{price\_range} & \textbf{capacity} \\
    \hline
    19/04/2024 & Amsterdam & Canal Stage & 25--70 EUR & 900 \\ \hline
    21/06/2024 & Seoul & Hanguk Arts & 30000--90000 KRW & 1100 \\ \hline
    10/12/2024& New York& The Metropolitan Opera& 80–280 USD& 3800\\ \hline
  \end{tabular}}
  \caption{d5: Event}
   \label{tab:Event}
  \end{subfigure}
  \hfill
    \begin{subfigure}[t]{0.51\textwidth}
    \centering
       \resizebox{\textwidth}{!}{ 
\begin{tabular}{|l|l|l|l|l|l|l|} \hline
\textbf{book\_title} & \textbf{author} & \textbf{format} & \textbf{price} & \textbf{currency} & \textbf{retailer} & \textbf{availability} \\ \hline
Carmen: A Novel & Prosper Mérimée & Paperback & 10.99 & USD & Amazon & In Stock \\ \hline
La Traviata & Alexandre Dumas fils & Hardcover & 18.50 & USD & Barnes \& Noble & In Stock \\ \hline
The Magic Flute & Emanuel Schikaneder & Paperback & 12.00 & USD & Bookshop.org & In Stock \\ \hline
Diane de Lys & Alexandre Dumas fils & Paperback & 14.50 & USD & Amazon & In Stock \\ \hline

\end{tabular}
}
    \caption{d6: Book}
    \label{tab:book}
  \end{subfigure}

  \begin{subfigure}[t]{0.7\textwidth}
    \centering
     \resizebox{\textwidth}{!}{ 
    \includegraphics[width=\linewidth]{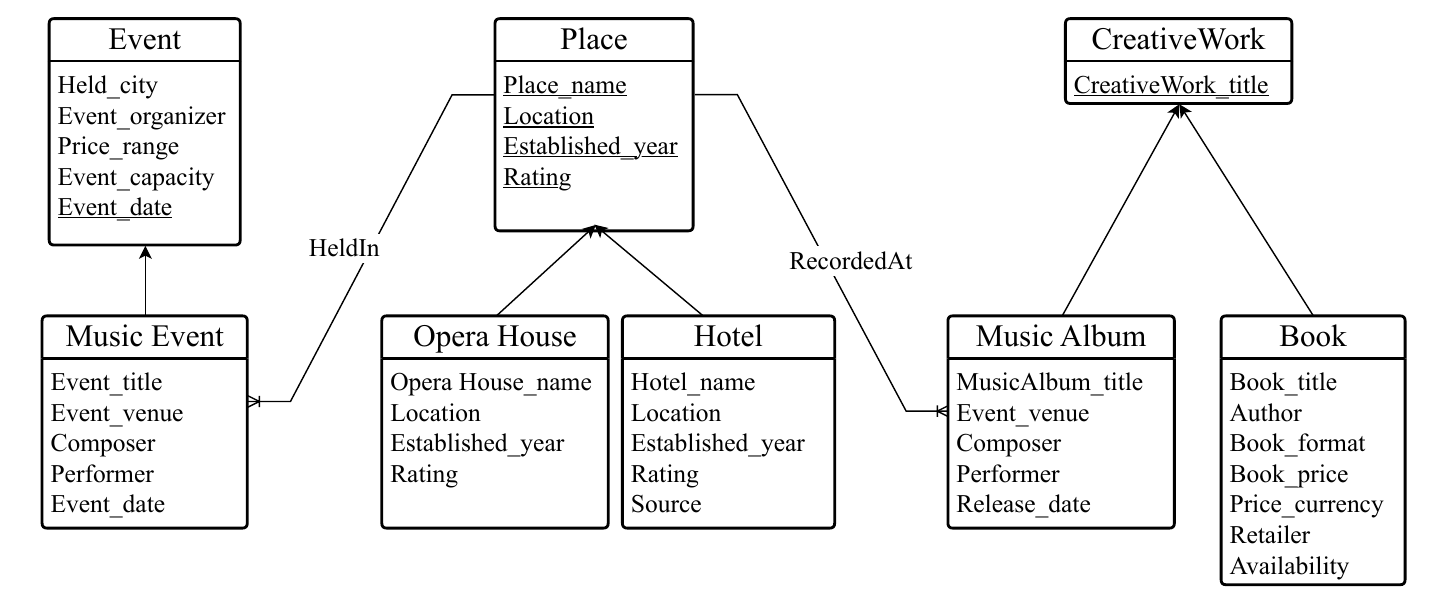}}
    \caption{Overall Conceptual Schema}
       \label{tab:ER}
  \end{subfigure}
  \caption{Example datasets and the inferred conceptual schema}
  \label{fig:ERtablesGraph}
\end{figure*}

{Consider the use case of dataset search, where Figure~\ref{tab:musicEvent}, which describes \textit{music events}, is the query table. 
The goal is to find tables describing related events and their venues. 
When using table-based retrieval, the system typically returns Figure~\ref{tab:musicRecording} (\textit{music recordings}), since its attributes (\textit{composer, performer, venue, release\_date}) and column semantics closely resemble those of the query table, even sharing overlapping values such as concert and opera names. 
It may also rank Figure~\ref{tab:book} (\textit{books about operas}) highly, because of partial overlaps in attribute domains (\textit{title} and \textit{author}) and value co-occurrences with composers’ names. 
However, Figure~\ref{tab:Event}, which also represents event-related data, is often missed, as its attributes (\textit{date, start\_city, organizer, price\_range, capacity}) share little domain overlap with the query table. 
This illustrates that relying solely on attribute-level overlap or domain alignment of columns cannot reliably determine whether two tables belong to the same semantic domain. 
In contrast, when searching with the keyword “opera”, the system may return tables about books or hotels (Figures~\ref{tab:book} and~\ref{tab:Hotel}) containing the word “opera”, while missing those that describe opera performances or music recordings.}

{
Prior work has attempted to address such issues by summarizing or organizing the structure of heterogeneous datasets. Some methods, often referred to as schema inference~\cite{DBLP:journals/vldb/CebiricGKKMTZ19,Kellou-Menouer-22,DBLP:conf/vldb/YuJ06a}, target semi-structured data such as XML, JSON, or RDF and rely mainly on syntactic similarity, making them unsuitable for inferring schemas from raw tabular data.
Similarly, techniques such as data profiling~\cite{Abedjan-15} and join discovery~\cite{ACM:proceedings:JOSIESigmod19,DBLP:conf/icde/DongT0O21} aim to enrich the logical schema by uncovering relationships between tables. The logical schema captures a repository’s raw structure — its tables and attribute definitions, but these techniques often fail to reduce schema complexity or resolve inconsistencies in representation.}

In contrast, we tackle the problem of conceptual schema inference from tables.
{A conceptual schema
provides a higher-level, semantically coherent view of the
repository, following the tradition of conceptual modelling
exemplified by the Entity-Relationship model~\cite{chenERD}.} 
It structures data into entity types (often organized in is-a hierarchies), their attributes, and relationships between types.
{For instance, given the tables in Figure~\ref{fig:ERtablesGraph}, a conceptual schema would identify types such as \textit{Event}, \textit{MusicEvent}, and \textit{Place}, together with their subtype relations (e.g., \textit{MusicEvent}~$\subset$~\textit{Event}). 
It would further associate attributes and relationships with these types; for example, \textit{MusicEvent} has the attribute \textit{venue}, and the relationship \textit{heldAt} towards \textit{Place}.
Such a schema offers a unified, explicit and conceptual view of the repository, allowing users to uncover semantically related tables, like those describing \textit{music events} and their \textit{venues}, that remain disconnected under existing search paradigms.}
Conceptual schema inference can complement dataset search by revealing which kinds of data are present, repository navigation by exposing recurring relationships, and annotation by grouping semantically related datasets and attributes even in the absence of a suitable ontology.

Conceptual schema inference from arbitrary tables is challenging especially when the repository is large and heterogeneous. For generality, we assume access to only minimal source metadata, such as table and column names.
Even with source metadata, the difficulty persists due to the following two key challenges.
The first challenge is semantic heterogeneity; for example, the same attribute can appear under different representations (e.g., \textit{location} in {Figure}~\ref{tab:Hotel} vs. \textit{start\_city} in {Figure}~\ref{tab:Event}), making it difficult to align semantically similar columns across tables.
The second is granularity variation, where tables mix records of entity types from different hierarchical levels; {for example, a table may mix records of cities, provinces, and special administrative regions within the same table.} 
Addressing these challenges calls for methods that can align semantic signals across datasets and generalize them into coherent conceptual type hierarchies and inter-type relationships.

Unlike prior work that relies on curated taxonomies to resolve inconsistencies~\cite{conf/sigmod23/santos}, we infer conceptual hierarchies directly from data, without external ontologies and without task-specific training. We develop two {approaches}, \methodG and \methodE, leveraging the strengths of decoder- and encoder-based large language models (LLMs), respectively. 
{\methodG infers local type hierarchies from table-level instances, identifies attributes from table columns, and derives semantic relationships across types by analyzing correspondences between attributes and their cell values. The inferred components are then aligned and integrated into a coherent global schema.}
\methodE relies on column-level semantic similarity captured by embedding. It clusters {tables/columns} in the embedding space to infer types and attribute groups, and infers relationships among types from {attribute level} similarities to construct the overall schema.

In summary, this work explores conceptual schema inference for tabular datasets. Our main contributions are as follows:
\begin{itemize}
\item {We formally define the problem of conceptual schema inference from tabular data, which aims to recover entity types, type hierarchies, attributes, and semantic relationships using only column names and cell values.}
   \item {Based on this definition, we propose two approaches for conceptual schema inference: \methodG, a generative LLM-based method 
   and \methodE, an encoder-based table embedding method, both capable of identifying entity types and hierarchies, inferring their attributes and discovering inter-type relationships.}
   
    \item {We conduct a systematic evaluation of both methods on real-world datasets, demonstrating their effectiveness and scalability to large repositories.} 

    \item We provide an end-to-end case study comparing the schemas inferred by the two methods, visualizing their resulting conceptual schemas to illustrate their respective strengths and limitations.  
\end{itemize}


This work dramatically extends our two previous papers \cite{wu2025taxonomyinferencet,wu2025schema}.
The earlier work \cite{wu2025taxonomyinferencet} deals with only type and type hierarchy inference from raw tables, while this work extends the problem to conceptual schema inference which includes two more sub-tasks, and improves the type and type hierarchy inference module in \cite{wu2025taxonomyinferencet} for more promising performance as demonstrated in our experiments.  
Compared to the more recent work \cite{wu2025schema} which is also to address the three sub-tasks of conceptual schema inference by proposing a generative LLM-based method SI-LLM, this work (1) extends SI-LLM to \methodG with a new method for inferring conceptual attributes, outperforming the original version, and a new module for inferring relationship cardinality, (2) presents a new column embedding-based method \methodE using encoder-based LLM, and (3) conducts more systematic evaluation with two more datasets.

The paper is organized as follows.
Section \ref{sec:related} reviews related work.
Section \ref{sec:problemd} formally defines the problem.
Sections \ref{sec:gesi} and \ref{sec:emsi} present the overviews and technical details of \methodG and \methodE, respectively.
Section \ref{sec:exp} reports our experimental evaluation, and Section \ref{sec:case} visualizes and compares the conceptual schemas inferred by the two methods.
Finally, Section \ref{sec:conclu} discusses possible extensions and concludes the paper.

\section{Related Work}
\label{sec:related}
We review related work on {\it schema inference} and {\it column similairty} which plays an important role in \methodE.

{\subsection{Schema Inference}}
{Schema inference }has been employed for different purposes by a variety of communities. 
Most research on schema inference has been closely associated with a specific data model, and features of the model {determine the intended purpose of inference}. 
For example, schema inference for XML~\cite{DBLP:conf/vldb/BexNV07,DBLP:journals/tods/BexNSV10} and JSON~\cite{DBLP:journals/vldb/BaaziziCGS19,DBLP:conf/iri/FrozzaMC18} tends to infer a schema that can precisely describe a collection of documents, with a view to ensuring that every document conforms to the inferred schema. This approach is best suited to settings where the schema is being inferred for a collection of largely consistent documents.  In contrast, schema inference for RDF~\cite{DBLP:journals/tlsdkcs/ChristodoulouPF15,DBLP:conf/er/Kellou-MenouerK15,DBLP:journals/vldb/GoasdoueGM20} tends to assume that the published data may be both diverse and from a variety of inconsistent publishers, and thus that the inferred schema is more summarising the contents of a repository than inferring a description to which all instances must conform.  This ethos is generalised by constructing a graph over different data models in \textsc{Abstra}~\cite{DBLP:conf/edbt/BarretMU24}, where the graph is then partitioned into collections that represent recurring structural patterns.  In XML and JSON, the primary evidence informing the inference process is the names of elements in documents; in RDF, documents may also be annotated with type information, and proposals differ in the extent to which they assume the use of consistent terminology. Nextia$_{DI}$ ~\cite{DBLP:journals/semweb/FloresRNGRJD24} also uses a graph model, though with a focus on the merging of schemas, on the basis of off-the-shelf pairwise similarity measures.

In contrast with approaches that infer schemas to which the original data must conform, several proposals aim to infer concise summaries that highlight the primary concepts represented in the data sets~\cite{DBLP:conf/vldb/YuJ06a,DBLP:journals/pvldb/YangPS09}. Such approaches may take as input a target size for the inferred schema, in which case the objective is then to identify the most representative datasets, which may be those that are extensively connected to others.  Although not explicitly on schema inference, ALITE supports the generation of a {\it full disjunction} of a collection of underlying tables~\cite{DBLP:journals/pvldb/KhatiwadaSGM22}.  In essence, ALITE identifies the set of equivalent columns across a collection of tables, and then derives the {\it full disjunction}; this involves 
taking the outer union, and then looking for maximally integrated tuples. Thus ALITE generates and populates an integration schema for a collection of tables. ALITE therefore shares with this paper the goal of producing a schema that spans multiple underlying tables. However, the approach differs from ours in mapping all datasets to a single table for integration, whereas we infer a much more granular schema. 

A related line of work applies LLMs to domain modeling from textual descriptions rather than tabular data ~\cite{DBLP:conf/er/ProkopSSKN24,DBLP:conf/er/SilvaMCKP24}. 
Recent studies infer type signals directly from tables without predefined ontologies~\cite{WangTRL2024DynoClass} and organize tables hierarchically for exploratory discovery~\cite{Fan2025HierTabSemantics}. 
While these methods recover useful type and hierarchical structure cues, they do not construct a repository-level conceptual schema that jointly models entity type hierarchies, attributes, and inter-type relationships while operating under minimal metadata assumptions.

\vspace{0.1cm}
\noindent
\subsection{Column Similarity} 
Column similarity surfaces in a variety of data integration tasks including schema matching, join discovery and dataset search.

In \textit{schema matching}, there has been a long history of work that has sought to identify related schema elements to inform data integration, with early work surveyed by Rahm and Bernstein~\cite{DBLP:journals/vldb/RahmB01}. More recently, Valentine~\cite{DBLP:conf/icde/KoutrasSIPBFLBK21} provided an experimental framework for the comparison of schema matching techniques, and compared several well established schema matching proposals with more recent embedding-based proposals, with results showing that established methods often competed effectively with {early} embedding based proposals.
{
Recent work has also explored LLMs for schema matching~\cite{DBLP:conference/kdd/Liu24,DBLP:journals/pvldb/LiuPSWF25}. 
Magneto~\cite{DBLP:journals/pvldb/LiuPSWF25} couples both small and large language models. In contrast, we focus on inferring repository-level conceptual structures directly, evaluating embedding-based and generative methods as alternative inference strategies rather than a joint pipeline.}
In experiments, we consider Unicorn~\cite{DBLP:journals/sigmod/FanTLWDJGT24}, which supports seven matching tasks, including schema matching and column type annotation, and learns classifiers for matching over embeddings that are designed to support different matching tasks.  Unicorn has been included as an embedding backbone because it is distinctive in combining knowledge from matching representations, and has been shown to perform well in comparison with a wide variety of state-of-the-art matching proposals~\cite{DBLP:journals/sigmod/FanTLWDJGT24}.

In \textit{join discovery}, the focus for similarity tends to be on overlaps in column values, on the assumption that the usefulness of a discovered join can be measured using a metric that relates to result size.  Join discovery may address equi-joins in which case column values need to match exactly (e.g., JOSIE~\cite{ACM:proceedings:JOSIESigmod19} or LSH-Ensemble~\cite{DBLP:journals/pvldb/ZhuNPM16}), or semantic-joins in which case column values need to satisfy some similarity measure (e.g., Aurum~\cite{DBLP:conf/icde/FernandezAKYMS18}, PEXESO~\cite{DBLP:conf/icde/DongT0O21}). In experiments, we consider the semantic-join version of DeepJoin~\cite{DBLP:journals/pvldb/Dong0NEO23} as an example of a join discovery proposal.  DeepJoin fine-tunes an encoder-based LLM for joinability, and is trained using joinable columns identified using PEXESO~\cite{DBLP:conf/icde/DongT0O21} as positive examples and random sampling of non-positively paired columns to produce negative examples. DeepJoin has been included as an {embedding backbone} because it is trained without manually labelled examples (and thus is self-supervised) and has been shown to perform well in comparison with a variety of state-of-the-art proposals, including LSH-Ensemble~\cite{DBLP:journals/pvldb/ZhuNPM16}.

In \textit{dataset search}, a common paradigm is \textit{table union search} (TUS), which searches for tables that can be unioned with a given table. In practice, this tends to involve looking for similar columns between a query table and tables in the repository \cite{DBLP:journals/pvldb/NargesianZPM18}. 
Proposals for TUS include two common solutions: (1) making use of indexes of table properties as evaluated by Taha \textit{et al.}~\cite{DBLP:conf/semco/TahaLSI24} and as implemented in D3L~\cite{DBLP:conf/icde/BogatuFP020} which uses a variety of extensional and intensional data features, and (2) relying either on encoder-based LLMs to capture similarity  \cite{DBLP:journals/pvldb/FanWLZM23,DBLP:conf/acl/HuWQLSFKKWY23} or {on} external information such as ontologies \cite{DBLP:journals/pvldb/Dong0NEO23}.
In experiments, we consider Starmie~\cite{DBLP:journals/pvldb/FanWLZM23} as an embedding backbone, representing the dataset search proposals that are built on column similarity.  
It uses a contrastive learning algorithm SimCLR~\cite{DBLP:conf/icml/ChenK0H20} to fine-tune a language model for column representations for comparison, and has been shown to perform better in comparison with a variety of dataset search proposals.


 \section{Problem Formulation}
\label{sec:problemd}
In this section, we define the problem of conceptual schema inference and some key concepts it involves. 
\begin{definition}[Conceptual Schema Inference]
Formally, let \(D=\{d_1,\ldots,d_n\}\) be a collection of tables. Each table \(d\in D\) has columns \(C(d)=\{c_1,\ldots,c_m\}\); each column \(c\) has a header and cell values \(V(c)=\{v_1,\ldots,v_k\}\). The task is to infer a conceptual schema \(S=\langle T,H,R\rangle\) that captures and unifies the structural properties of \(D\), where 
\(T\) is the set of conceptual types with related attributes, \(H\) is the type hierarchy, and \(R\) is a set of directed relationships between types in \(T\).
\end{definition}

\begin{definition}[Conceptual Type]
\(T\) is a set of {conceptual types}. Each \(t\in T\) has a name \(t.\mathit{typeName}\) and a set of attributes \(\text{Atts}(t)=\{a_1,\ldots,a_p\}\). Each type \(t\) is also linked to a subset of tables from which it is derived, \(t.\mathit{datasets}\subseteq D\).  Intuitively, a \textit{conceptual type} captures the shared semantic domain across one or more tables. That is, if the rows in a table represent instances of a coherent real world concept (e.g., company or financial transaction), these records can be viewed as instances of the same conceptual type. We use \textit{conceptual type} and \textit{entity type} as synonyms in what follows, and refer to them simply as \textit{types} when no ambiguity arises.
\end{definition}

\begin{definition}[Type Hierarchy] {A type hierarchy is a directed acyclic graph $H=(T,E)$, where each node corresponds to a type in $T$, and each edge $(u,v)\in E$ with $u \in T$ and $v \in T$ indicates that $u$ is a subtype of $v$.}
In this work, we construct hierarchies that are rooted at a built-in top type \textit{Thing}\footnote{Note we may omit \textit{Thing} for convenience in presenting example hierarchies in the paper.}.

\end{definition}

 \begin{definition}[Top-Level Type]
 {
 Given a type hierarchy $H=(T,E)$ rooted at \textit{Thing}, a \emph{top-level type} is a direct subtype of \textit{Thing}:
$T_{\text{top}} = \{\, t \in T \mid (\,t,\textit{Thing}\,) \in E \,\}.
$
These types have no supertypes except for \textit{Thing}, representing the most general concepts in the schema.
Their descendants form a layered structure capturing increasingly specific \emph{is-a} relationships.}
\end{definition}


\begin{definition}[Conceptual Attribute]
Given a type \(t \in T\), a conceptual attribute \(a \in \texttt{Atts}(t)\) is a property of \(t\)
that takes its values from a semantic domain \(Dom(a)\) and plays a specific role in describing some aspect of \(t\). {\(Dom(a)\) can be either a primitive data type (e.g., \textit{String}, \textit{Number}, \textit{Date}, \textit{Time}, \textit{Identifier}) or a conceptual type (e.g., \textit{Person}, \textit{Organization}).}
Each conceptual attribute corresponds to a group of columns,
denoted \(a.{datasetColumns} \subseteq \bigcup_{d \in D} C(d)\). {For each table \(d\), the set of conceptual attributes corresponding to its columns is denoted by \(d.{conceptualAttributes}\).}
Conceptual attributes that share the same semantic domain are treated as distinct when serving different roles. 
For example, in the \textit{Music Recording} type, attributes \textit{Composer} and \textit{Performer} both refer to persons but denote different roles within the type. We refer to conceptual attributes simply as \textit{attributes} throughout the paper
\end{definition}

\begin{definition}[Relationship]
Each relationship $r \in R$ is a quadruple $\langle t_i,a_i,t_j, p\rangle$, where $t_i,t_j\in T$, $a_i\in \texttt{Atts}(t_i)$ is an attribute of $t_i$ whose values denote instances of type $t_j$, and $p$ is a semantic predicate that {expresses the meaning of the relationship between} $t_i$ and $t_j$ via $a_i$.
\end{definition}

\begin{definition}[Relationship Cardinality]
Given a relationship \(r = \langle t_i, a_i, t_j, p\rangle \in R\),
its cardinality, denoted as \(Card(t_i, t_j)\), is a function that describes the mapping multiplicity between instances of \(t_i\) and \(t_j\), indicating the number of instances of one type that can be associated with those of the other.
This work distinguishes four canonical cardinality patterns, namely:
one-to-one (1:1), one-to-many (1:n), many-to-one (n:1), and many-to-many (m:n).
\end{definition}

 \section{Schema Inference with Generative LLMs 
 }
\label{sec:gesi}

\begin{figure*}[tb]
    \centering
          \resizebox{\textwidth}{!}{ 
    \includegraphics[width=\linewidth]{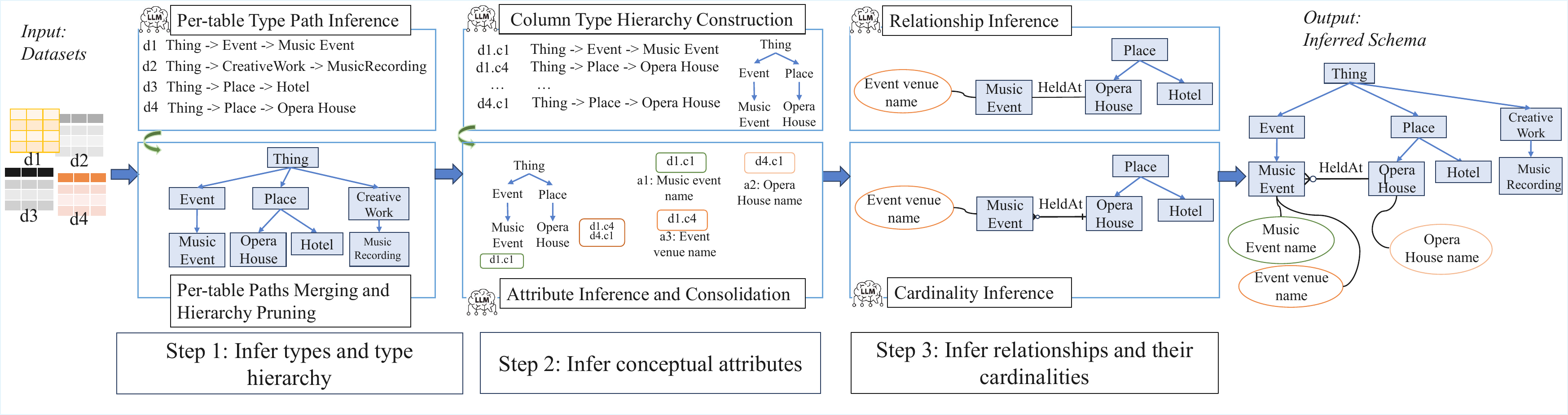}
    }
    \caption{The overall framework of \methodG}
    \label{fig:gesiOverall}
\end{figure*}

{Repository-level conceptual schema inference with generative LLMs raises three main challenges. 
First,  schema inference requires decomposition, since prompting an LLM with all tables and columns is generally impractical and difficult to control, yet local outputs must still be consolidated into a coherent global schema. 
Second, the inferred schema must manage conceptual granularity, avoiding types driven by individual instances or overly broad types that collapse distinct concepts. 
Third, schema elements require contextual interpretation: attributes with similar value domains may play different roles under different types; relationships and cardinalities may rely on heterogeneous table evidence.}


{These challenges motivate the hierarchy-first design of \methodG. 
The design follows the dependencies among schema components: types provide context for attributes, while inferred types and attributes support relationship and cardinality inference. 
\methodG therefore constructs a schema through local-to-global steps, consolidating intermediate outputs for reuse in later stages. 
As shown in Figure~\ref{fig:gesiOverall}, \methodG proceeds through three sequential steps:
}
\begin{itemize}[leftmargin=*]
  \item Step 1: {\it Infer types and type hierarchy}. 
Given a set of tables $D$, \methodG infers table-specific type paths from individual tables.
{These paths are then merged into a repository-level hierarchy $H$ by aligning type names, consolidating equivalent types, and pruning inconsistent \textit{is-a} relationships.}
\item Step 2: {\it Infer conceptual attributes}. 
{Conditioned on the inferred type hierarchy, \methodG organises columns into a column-type hierarchy and resolves them into conceptual attributes. It then associates the resulting column groups with conceptual types and promotes shared attributes to appropriate parent types.}
\item Step 3: {\it Infer relationships and their cardinalities}. \\
Given a type and its conceptual attributes, \methodG {uses sampled attribute values to identify potential references to instances of other types. 
For each relationship, it then infers {the relationship predicate}, and determines the relationship cardinalities between type pairs by jointly considering the involved types, the relationship predicate, and contextual evidence from the related tables.}

\end{itemize}

The following subsections provide a detailed explanation to each step.

\subsection{Type and Type Hierarchy Inference}
\label{sec:typehierarchy}

In this step, we construct a global type hierarchy {$H = (T, E)$} over the table collection $D$, where each type $t\in T$ is associated with the tables that describe its instances. {We summarize the process in Algorithm~\ref{alg:type_hierarchy}.}


\subsubsection{Per-table Type {Path} Inference}
\label{sec:perTableHier}
\begin{figure*}[tb]
  \centering
  \begin{minipage}[b]{0.5\textwidth} 
  \vspace{0pt}
    \centering
   \includegraphics[width=0.85\linewidth]{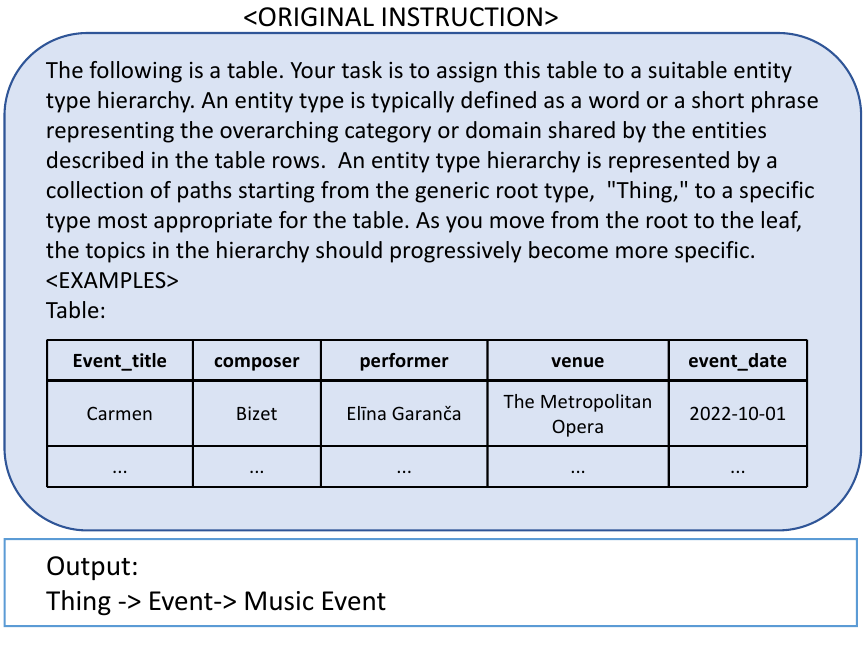}
    \caption{Basic prompt for per-table type path inference \protect\footnotemark}
    \label{fig:P1BasicPrompt}
 \end{minipage}
 \begin{minipage}[b]{0.48\textwidth}
 \vspace{0pt}
  \centering
    \includegraphics[width=0.8\linewidth]{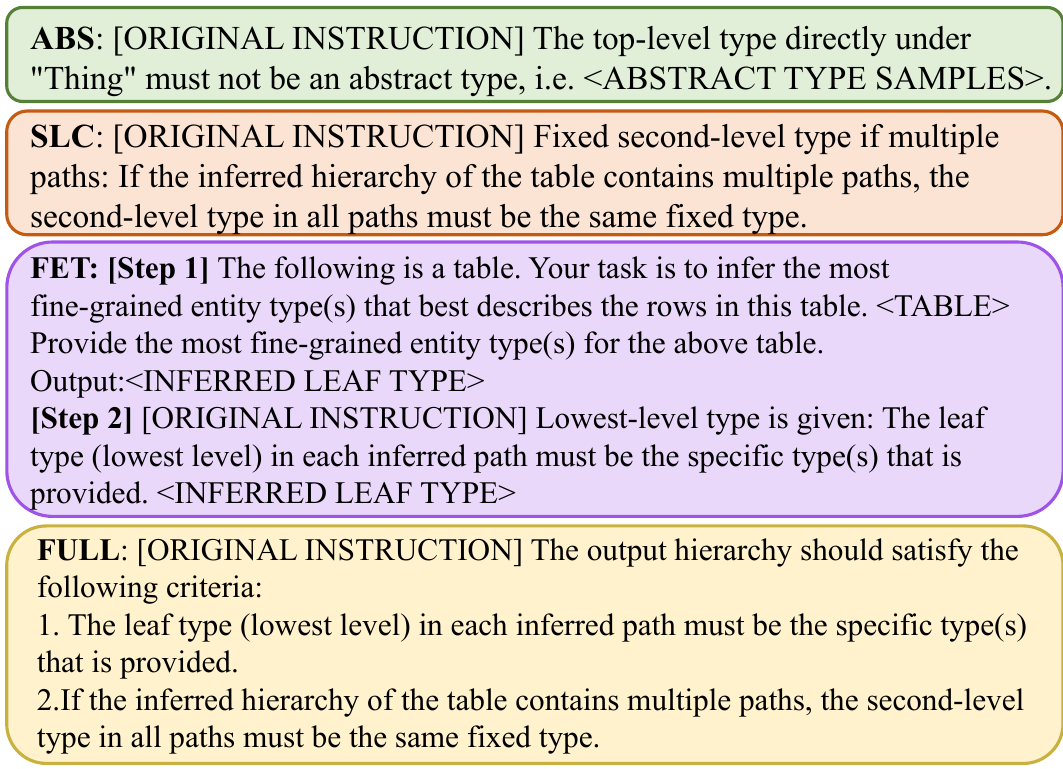}
    \caption{Prompt constraints for per-table type path inference}
    \label{fig:P1Prompts}
    \end{minipage}
\end{figure*}

We use the prompt shown in Figure~\ref{fig:P1BasicPrompt} to infer type {paths} for each individual table.
The input consists of sampled table records, and the output is type 
{paths} from the the root type {\textit{Thing}} to the most specific type of all the entities in the table with a maximum length of $N$ (e.g., \noindent{\ttfamily\footnotesize Thing → Place → Hotel}).
Each prompt includes three few-shot examples following \cite{DBLP:conf/vldb/kayaliChorus24}, where each example provides (i) a five-row table sample and (ii) its ground truth output type path(s). 
{The model is prompted to output one or multiple type paths per table.
Note that two paths are regarded as different if their leaves are the same but some intermediate types are different (e.g.,
\noindent{\ttfamily\footnotesize
Thing → CreativeWork → SoftwareApplication → VideoGame} is different from
\noindent{\ttfamily\footnotesize Thing → CreativeWork → Game → VideoGame}}). 

We extract all well-formed type paths from the LLM's output which includes extra content beyond the intended paths, using regular expressions. We retain the longest valid path
(\noindent{\ttfamily\footnotesize {Thing}→ $t_1$→$\cdots$→ $t_k$}, with each $t_i$ being an {inferred} type)
{and remove any trailing text that appears after detecting $N$ valid types in the path.} 
If duplicated types are detected within a path (e.g., repeated nodes such as \noindent{\ttfamily\footnotesize Thing → Product → Product}), the output is deemed malformed and the prompt is re-executed.

Notably, to construct the table input for each prompt, we apply a {context sampling} strategy to the input tables. Specifically, we randomly select 5 non-duplicate rows from each table to capture the semantic granularity across rows. 
Similarly, when columns serve as inputs in later stages of \methodG, up to 5 distinct values are randomly sampled from each column.
This sampling strategy is consistently adopted across all the prompts.

\footnotetext{
{To support reproducibility, we have made the implementation, prompts and datasets for both methods publicly available at \url{https://github.com/PierreWoL/SILM/}.
}}
\textit{Constraints in Prompts}. We further extend the original prompt in Figure~\ref{fig:P1BasicPrompt} by introducing constraints to ensure that inferred {per-table type paths} can be coherently merged into a global structure. These constraints aim to guide LLMs toward producing more structured and well-formed {paths}, and are treated as design parameters.
Motivated by recent work questioning the stability of prompt engineering \cite{sclar2024quantifying}, we systematically evaluate whether lightweight, interpretable constraints can enhance {path} inference.
We experiment with four extended prompts with constraints, as shown in Figure~\ref{fig:P1Prompts}, where the original prompt is denoted as \texttt{<ORIGINAL INSTRUCTION>}:
\begin{itemize}[leftmargin=*]
    \item \textit{Abstract Type Constraint (ABS):} This constraint avoids overly generic top-level types by blacklisting abstract types, {such as \textit{Object} and \textit{Collective Entity}}, which are direct subtypes of \textit{Entity} (Q35120) of Wikidata.
    \item \textit{Second-Layer Convergence Constraint (SLC):} This constraint promotes cross-table coherence by requiring that the top-level types converge across tables, discouraging their divergence.
    \item \textit{Most specific Type Constraint (FET)}: This constraint enforces two-step generation: first inferring the most specific types for a table, then generating paths that terminate at these types. 
    Note that the two steps are separately executed with the LLM, and the most specific types of each table are pre-extracted.
 
    \item \textit{Full Constraint (FULL):} This constraint combines FET and {SLC}.
\end{itemize}
The evaluation results (Section~\ref{sec:ablationStudy}) demonstrate that these constraints consistently improve type paths inference, each contributing in different ways.

\subsubsection{Per-table Paths Merging and Hierarchy Pruning}
\label{sec:mergePrune}
The final hierarchy is obtained by merging and pruning all per-table type {paths}. 
An initial merged hierarchy is first constructed by unifying all nodes and edges from individual {paths}. {Given \(H_1 = (T_1, E_1), \ldots, \\H_n = (T_n, E_n)\), we define
\(H_{\text{raw}} = (T_{\text{raw}}, E_{\text{raw}})\),
where \(T_{\text{raw}} = \bigcup_{i=1}^n T_i\) and \(E_{\text{raw}} = \bigcup_{i=1}^n E_i\).}
Each edge \((u,v) \in E_{\text{raw}}\) is assigned a weight
$w(u,v) = \sum_{i=1}^n \mathbf{1}\!\left[ (u,v) \in E_i \right],$
where \(\mathbf{1}[\cdot]\) is an indicator function that returns 1 if an edge is inside an edge set, and 0 otherwise.
Thus, \(w(u,v)\) represents the number of {path}s in which the edge \((u,v)\) appears.
An edge is assigned a higher weight if it appears in in more per-table type paths. 
%
We then apply the following pruning steps to \(H_{\text{raw}}\):
\begin{enumerate}[leftmargin=*]
    \item \textit{Self-loop pruning:} 
    Remove all self-loops \((u,u) \in E_{\text{raw}}\).
    
    \item \textit{Inverse-edge pruning:} 
    Convert bidirectional edges into unidirectional ones. 
    Specifically, if both \((u,v)\) and \((v,u)\) exist and \(w(v,u) > w(u,v)\), remove \((u,v)\).
    
    \item \textit{Erroneous \textit{is-a} relationship pruning:} 
    LLMs may occasionally introduce spurious or inconsistent \textit{is-a} relationships when generating per-table paths. These errors may propagate after merging. 
    For instance, the erroneous path \noindent{\ttfamily\footnotesize{Thing → Event → CreativeWork → MusicAlbum}} places \textit{CreativeWork}, which is a top-level type, under \textit{Event} as its subtype. 
    To mitigate such errors, we incorporate an LLM-based verification step: 
    for each candidate edge \((A,B)\), the LLM is prompted with 
    \textit{``Is A the parent type of B? Only answer yes or no.''} 
    Relationships with a response of no are removed. 
It is proposed that such a task-specific LLM-based verification manner can effectively reduce hallucination \cite{conf/neurips23/selfRefine}. {The contribution of this verification step is further assessed in an ablation study in Section~\ref{sec:ablationStudy}; the results show that removing it degrades hierarchy quality, confirming its role in reducing erroneous \textit{is-a} relationships.}
\end{enumerate}

In summary, \methodG derives type {paths} for each table and then merges them into a global hierarchy followed by pruning. We hypothesize that given sufficient table context, the LLM can leverage its pretrained knowledge to infer globally consistent type {paths} that concisely reflect the concepts of the entities in the tables. 

\begin{algorithm}[tb]
\caption{Infer types and their hierarchy in \methodG (Step 1).}
\label{alg:type_hierarchy}
\KwData{Set of dataset collection $D$}
\KwResult{Global type hierarchy $H = (T, E)$}
{Initialize weighted relation set $R \leftarrow \emptyset$\;
\For{$d \in D$}{
    $Answer_d \leftarrow$ \texttt{InferTableTypeHierarchy}$(d)$\; \label{line:pertypeh}
    $Paths_d \leftarrow$ \texttt{ExtractTypePaths}$(Answer_d)$\;\label{line:extractP}
    \ForEach{$(u, v)$ in $Paths_d$}{\label{line:mergeISA}
        $R[(u, v)] \mathrel{+}= 1$
    }
}
$H_{\text{raw}} \leftarrow$ \texttt{BuildHierarchy}$(R)$\; \label{line:buildGH}
$H \leftarrow$ \texttt{PruneHierarchy}$(H_{\text{raw}})$\;
\Return{$H = (T, E)$}\;}
\end{algorithm}

\subsection{Conceptual Attribute Inference}

\begin{figure*}[tb]
  \centering
  \begin{minipage}[b]{0.32\textwidth} 
    \centering
   \includegraphics[width=\linewidth]{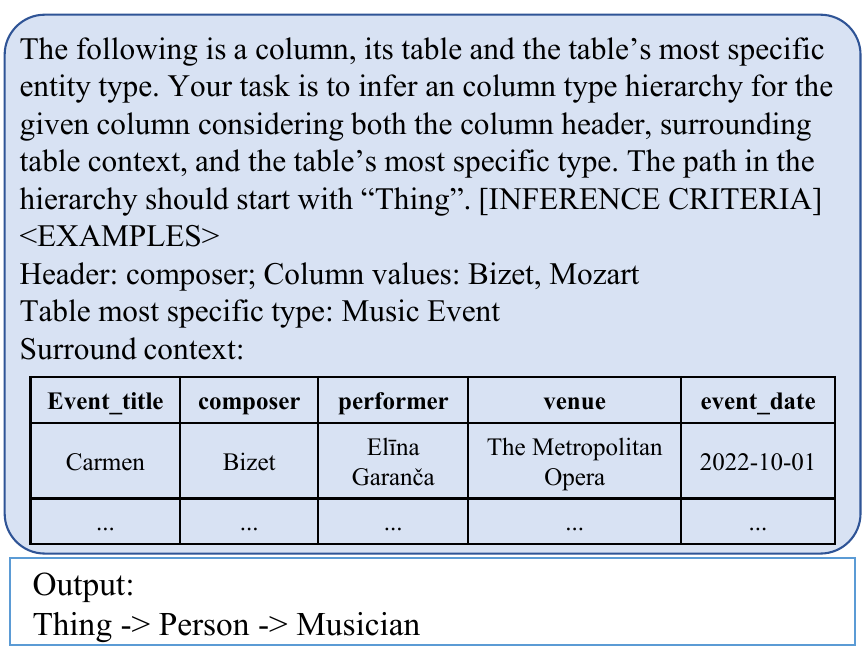}
      \caption{Prompt for inferring column type paths}
  \label{fig:cthp}
 \end{minipage}
 \begin{minipage}[b]{0.32\textwidth}
  \centering
    \includegraphics[width=\linewidth]{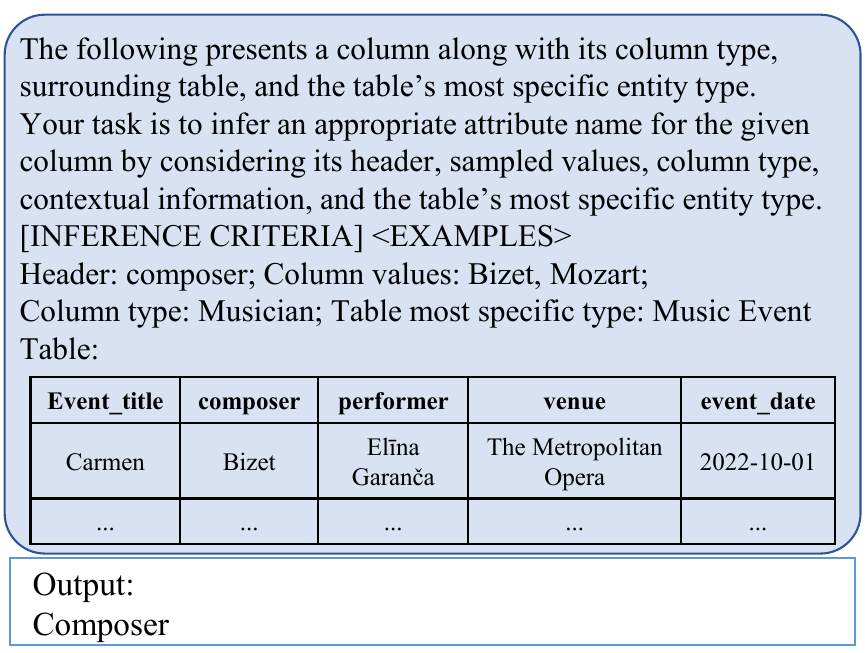}
    \caption{Prompt for inferring attribute names}
    \label{fig:ainp}
    \end{minipage}
    \begin{minipage}[b]{0.32\textwidth}
  \centering
    \includegraphics[width=\linewidth]{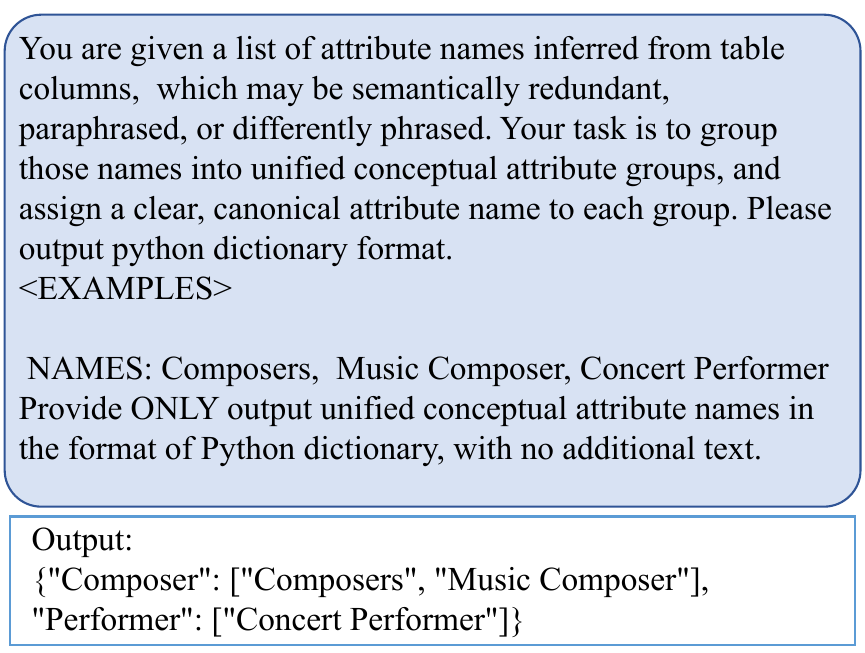}
    \caption{Prompt for resoluting attribute names}
    \label{fig:arp}
    \end{minipage}
\end{figure*}

The second step of \methodG is to infer conceptual attributes for each inferred type from the table columns. 
Inferring such attributes across large and heterogeneous tables is non-trivial. Current LLMs have limited context windows, making it infeasible to directly reason over an input composed of thousands of columns to identify shared attribute semantics. Naïve approaches such as direct LLM prompt-based grouping either fail to scale or produce inconsistent and overly generic groupings.

To address these challenges, we model attribute inference as a hierarchical reasoning process that integrates column-level semantics with table-level semantics, based on an intermediate column type hierarchy.
The procedure is summarized in Algorithm~\ref{alg:attribute-inference-coltype}. 
It proceeds in two phases: (i) construction of a column type hierarchy, and (ii) bottom-up attribute inference along the inferred table type hierarchy based on inheritance.

\begin{algorithm}[tb]
\caption{Infer conceptual attributes in \methodG (Step 2).}
\label{alg:attribute-inference-coltype}
\centering
\scalebox{0.9}{
\begin{minipage}{0.9\columnwidth}
\KwData{The dataset collection $D$, and the conceptual type hierarchy $H = (T, E)$ inferred from $D$}
\KwResult{Conceptual attributes $\texttt{Atts}(t)$ for each type $t \in T$}

\ForEach{$d \in D$}{
    \ForEach{$c \in d.columns$}{
        $\texttt{TypePath}_c \gets \texttt{InferColSemTypeHierarchy}(c, d)$\; \label{line:inferColType}
    }
}
$H_{\text{col}} \gets \texttt{MergeColTypeHierarchies}(\{\texttt{TypePath}_c \mid c \in \bigcup_{d \in D} d.columns\})$\; \label{line:mergeColType}

\tcp{Traverse hierarchy bottom-up for local and inheritance attribute inference}
\ForEach{$t \in H$ in bottom-up order \label{line:traversal}}{
    $D_t \gets \{ d \in D \mid \texttt{MostSpecificType}(d) = t \}$\;
    $C_{D_t} \gets$ all columns from tables in $D_t$\;

    \ForEach{$c \in C_{D_t}$}{
        $\texttt{SemType}(c) \gets \texttt{FindLCA}(c, H_{\text{col}})$\; \label{line:findLCA}
    }

$G^{t}_{\text{sem}} = \{\,(\tau,\ \{\,c \in C_{D_t}\mid \texttt{SemType}(c)=\tau\,\}) \mid \tau \in \{\texttt{SemType}(c)\mid c\in C_{D_t}\}\,\}$\;
 \label{line:colCluster}
    \ForEach{$g \in {{G}}^{t}_{\text{sem}}$}{
        \ForEach{$c \in g$}{
            $a_c \gets \texttt{InferAttributeName}(c, \texttt{SemType}(c), t)$\;\label{line:inferAttrN}
        }
        $\texttt{Atts}(g) \gets \texttt{UnifyAttributes}(\{a_c \mid c \in g\})$\;\label{line:AttrResolu}
    }

    $\texttt{Atts}(t) \gets \bigcup_{g \in {{G}}^{t}_{\text{sem}}} \texttt{Atts}(g)$\;\label{line:collectAttrs}

    $\texttt{Children}(t) \gets \{\, t' \in T \mid (t', t) \in E \,\}$\;\label{line:findCT}
    \If{$\texttt{Children}(t) \neq \emptyset$}{
        $\texttt{InheritedAtts}(t) \gets \texttt{UnifyAttributes}\!\left(
            \bigcup_{t' \in \texttt{Children}(t)} \texttt{Atts}(t')
        \right)$\;\label{line:inferInheritedA}

        \ForEach{attribute $a \in \texttt{InheritedAtts}(t)$}{
            \If{$\frac{|\{\, t' \mid a \in \texttt{Atts}(t') \,\}|}{|\texttt{Children}(t)|} \geq \theta$}{
                Add $a$ to $\texttt{Atts}(t)$\;\label{line:addInheritedA}
            }
        }
    }
}
\Return{$\{\,\texttt{Atts}(t) \mid t \in T\,\}$}
\end{minipage}
}
\end{algorithm}

\subsubsection{Column Type Hierarchy Construction}
\label{sec:cthc}
We construct a global column type hierarchy \(H_{\text{col}}\) by inferring, merging and pruning the type paths from individual columns in the whole table collection $D$, so as to capture the shared column semantics. The inference process follows Type and Type Hierarchy Inference of Step 1 of \methodG (cf. Section~\ref{sec:typehierarchy}), with two main differences. First, we use a prompt tailored for per-column type path inference (Line~\ref{line:inferColType}), where the input describes the target column through five sampled cell values together with the rows containing them, providing the surrounding table context, as shown in Figure~\ref{fig:cthp}. Second, we omit the LLM-based verification in the hierarchy pruning stage, given the large number of type paths that need be merged during hierarchy construction.


\subsubsection{Attribute Inference and Consolidation}
\label{sec:attrInherit}
For each type \(t\) in the table type hierarchy \(H\), we collect all the tables whose most specific inferred type is \(t\). 
Each column in these tables is then mapped to the {lowest common ancestor} (LCA) of its candidate column type paths within \(H_{\text{col}}\) {(Line~\ref{line:findLCA})}. 
This mapping anchors each column to a correct but fine-grained enough type.
For example, a column that contains both individual musicians and music bands, which belong to the branches of the top-level types of \textit{Person} and \textit{Organization}, is mapped to the LCA \textit{Artist}. 

Columns that are mapped to the same type in the column type hierarchy are grouped into one semantic column cluster {(Line~\ref{line:colCluster})}, representing a potential attribute domain for type \(t\). 
Within each cluster, the LLM is prompted to infer an attribute name for each column, reflecting its semantic role within the table while considering its column name, sampled values, column type and table-level context ({Line~\ref{line:inferAttrN}-\ref{line:AttrResolu})}. See Figure~\ref{fig:ainp} for this prompt for attribute name inference.
Note that columns of the same type may serve distinct roles in different tables. 
For instance, for two columns of the type \textit{Person} in a table describing music recordings, one may represent the \textit{Composer} and the other the \textit{Performer}. 

Besides, to ensure naming consistency across semantically equivalent attributes, an additional LLM-based resolution step is applied within each column-type group to consolidate synonymous names. The prompt is shown in Figure~\ref{fig:arp}. 
The unified attributes across all column-type groups collectively form the local attribute set \(\texttt{Atts}(t)\) for type \(t\).

Finally, the algorithm performs attribute inheritance inference and consolidation, implemented through bottom-up traversal of the table type hierarchy
{(Line~\ref{line:findCT}-\ref{line:addInheritedA})}. 
For a type $t$, attributes shared among its direct subtypes are considered as its attributes. 
In particular, if an attribute appears in at least a proportion $\theta$ of the direct subtypes, it is promoted to $t$ as its inherited attribute. 
This proportion-based strategy enables the attribute inference through type inheritance, but also preserves the specialization of each subtype. 

\subsection{Relationship and {Relationship} Cardinality Inference}

\begin{figure*}[tb]
  \centering
  \begin{minipage}[b]{0.32\textwidth} 
    \centering
   \includegraphics[width=\linewidth]{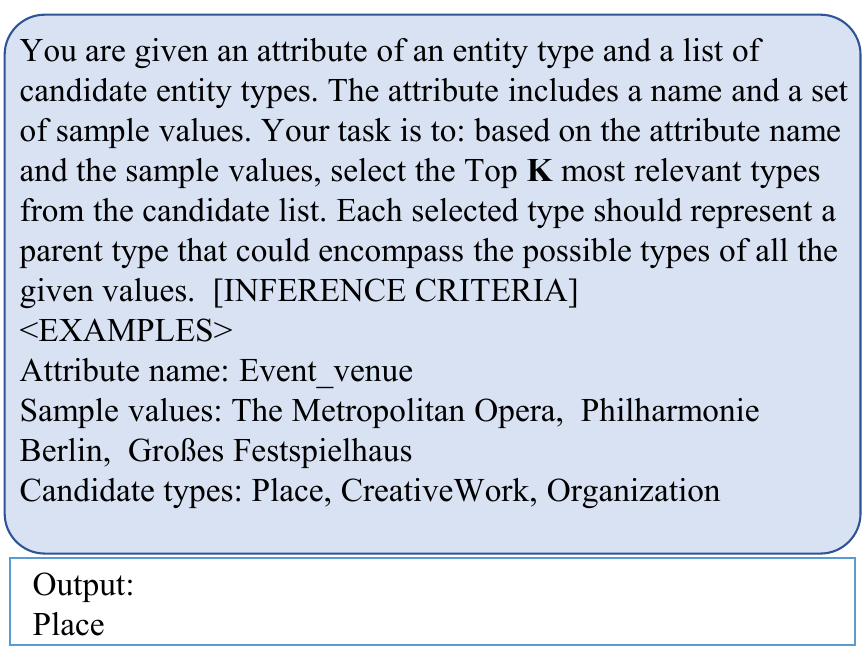}
      \caption{Prompt for selecting top-$K$ candidate top-level types in relationship inference}
  \label{fig:rdd}
 \end{minipage}
 \begin{minipage}[b]{0.32\textwidth}
  \centering
    \includegraphics[width=\linewidth]{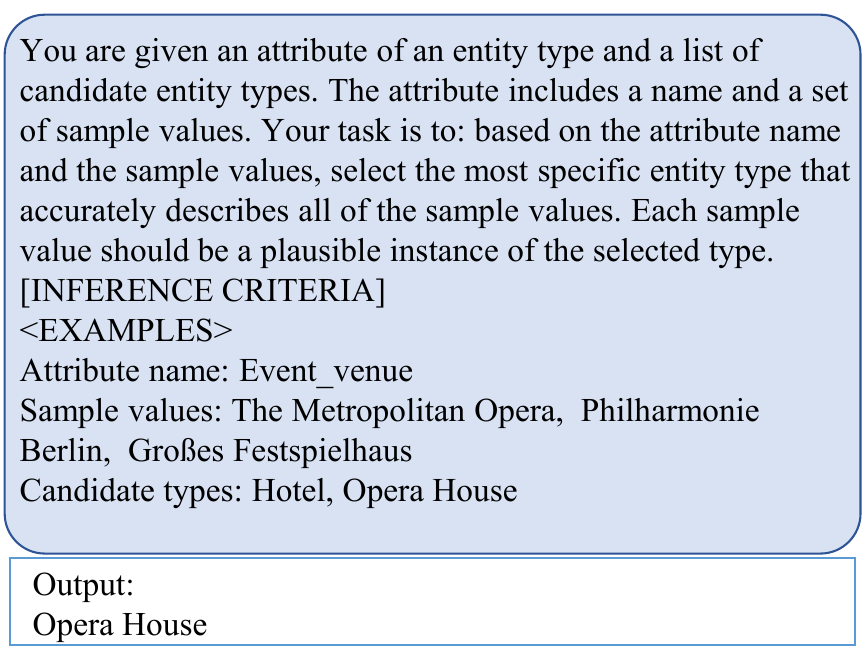}
    \caption{Prompt for selecting the most plausible specific type in relationship inference}
    \label{fig:rds}
    \end{minipage}
    \begin{minipage}[b]{0.32\textwidth}
  \centering
    \includegraphics[width=\linewidth]{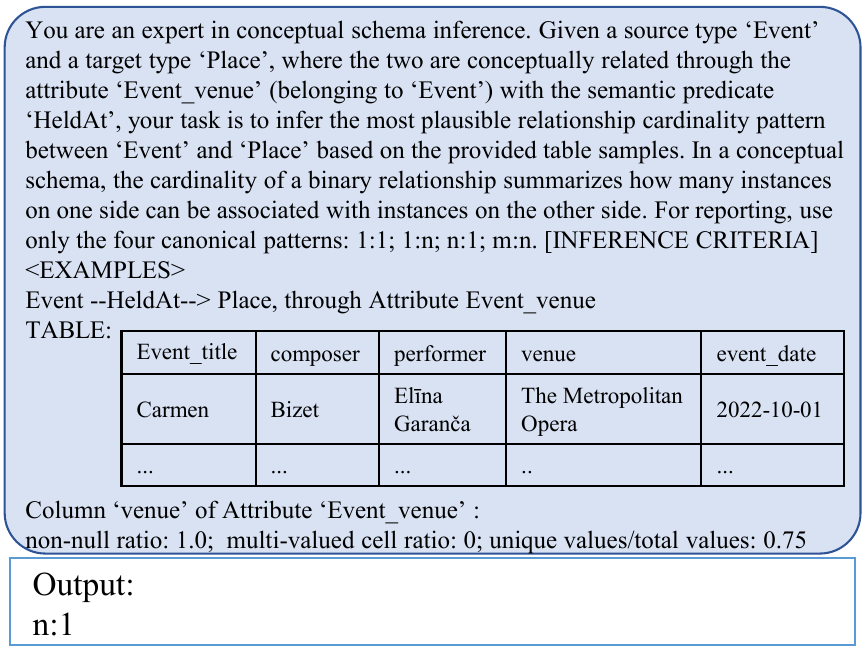}
    \caption{Prompt for relationship cardinality inference for an example relationship of HeldAt}
    \label{fig:rcd}
    \end{minipage}
\end{figure*}

\subsubsection{Relationship Inference}\label{sec:relationI}
We infer relationships that are implicitly indicated by attribute values that refer to instances of other types. Based on the definition in Section~\ref{sec:problemd}, we identify such relationships by detecting conceptual attributes whose values are named entities corresponding to instances of another inferred type.
For example, consider the type \textit{Music Event} which has the attribute \textit{Event\_venue}, {with} values, such as “\textit{The Metropolitan Opera}”, “\textit{Teatro alla Scala}”, “\textit{Philharmonie Berlin}”, recognizable instances of the type \textit{Place}.
This attribute \textit{Event\_venue} indicates a semantic relationship from \textit{Music Event} to \textit{Place}.


Algorithm~\ref{alg:relation_inferenceTopK} outlines the procedure of \methodG for relationship inference.
For each target type $t_i$, \methodG identifies conceptual attributes whose columns contain named entities. 
Following the same context sampling strategy used in per-table type path inference (cf. Section~\ref{sec:perTableHier}), each conceptual attribute $a_i$ is represented using {$5$} sampled cell values from its source columns (Line~\ref{line:samplecol}).
For each such attribute $a_i$, a top-down type hierarchy traversal strategy is then applied to determine the target type $t_j$ that the attribute values most likely belong to, thereby establishing a relationship between $t_i$ and $t_j$.
Specifically, the LLM first selects the top-$K$ candidate top-level types $T'_{\text{top}}$ from the table type hierarchy (Line~\ref{line:topktype}), using the prompt shown in Figure~\ref{fig:rdd}. These types are potential types of the sampled instances. 
The search is then refined within the union of descendants of $T'_{\text{top}}$, where the LLM identifies the most matched type $t_j$ for $a_i$ (Line~\ref{line:specifictype}, prompt in Figure~\ref{fig:rds}).

Once the target type $t_j$ is determined, an LLM generates a predicate label $p$ for the relationship, taking as input the triplet $\langle t_i, a_i, t_j \rangle$ (Line~\ref{line:predicate}). 
As illustrated in Figure~\ref{fig:gesiOverall}, for the conceptual attribute \textit{Event\_venue} of the type \textit{Music Event}, \methodG first identifies that the attribute values correspond to the top-level type \textit{Place}, and then refines this type to its descendant \textit{Opera House}. 
Finally, the LLM-inferred predicate \textit{HeldAt} denotes a relationship which means a \textit{Music Event} takes place in an \textit{Opera House}.

\begin{algorithm}[tb]
\caption{Infer relationships between types in \methodG (Step 3).}
\label{alg:relation_inferenceTopK}
\centering
\scalebox{0.9}{
\begin{minipage}{\columnwidth}
\KwData{The inferred table type hierarchy $H=(T,E)$; 
a target type $t_i$ with associated conceptual attributes $\texttt{Atts}(t_i)$ and columns $\texttt{C}_{t_i}$;
the set of type-level types $T_{\text{top}}$ of $H$ as the candidates; the number $K$ of retained top-level type candidates used in \texttt{TopKMatchTopLevelTypes} for inferring relationships from $t_i$ to other types in $H$.
}
\KwResult{A set of relationships $\texttt{R}(t_i)$ between $t_i$ and other types in $H$}
$\texttt{R}(t_i) \gets \emptyset$\;

$\texttt{Atts}_{\text{NE}}(t_i) \gets$ Filter attributes in $\texttt{Atts}(t_i)$ whose columns contain named entities\;

\ForEach{$a_i \in \texttt{Atts}_{\text{NE}}(t_i)$}{
  $V_{a_i} \gets \texttt{Sample}\!\left(\bigcup a_i.\texttt{datasetColumns}, N\right)$\;\label{line:samplecol}
  $T^{\text{top}}_K \gets \texttt{TopKMatchTopLevelTypes}(a_i, V_{a_i}, T_{\text{top}}, K)$\;\label{line:topktype}

  \ForEach{$t_{\text{top}} \in T^{\text{top}}_K$}{
    $T^{\text{desc}} \gets \texttt{GetDescendantTypes}(H, t_{\text{top}})$\;
    $t_j \gets \texttt{SelectMostSpecificType}(a, V_{a_i}, T^{\text{desc}})$\; \label{line:specifictype}

    \If{$t_j \neq \text{NULL}$}{
      $p \gets \texttt{InferRelationshipPredicate}(t_i, a_i,t_j)$\; \label{line:predicate}
      Add $\langle t_i, a_i,t_j, p \rangle$ to $\texttt{R}(t_i)$\;
    }
  }
}
\Return{$\texttt{R}(t_i)$}
\end{minipage}
}
\end{algorithm}

\subsubsection{Cardinality Inference}

Once we infer a relationship $\langle t_i, a_i, t_j, p \rangle$ together with a predicate label $p$
, the next step is to determine its most plausible cardinality.
To this end, we provide the LLM with statistical evidence derived from the columns associated with the attribute $a_i$, such as value distinctness and frequency, together with contextual information of their tables.
The overall process {of} inferring cardinality is shown in Algorithm~\ref{alg:cardinality_inference}.

\begin{algorithm}[tb]
\caption{Infer relationship cardinality in \methodG (Step 3).}
\label{alg:cardinality_inference}
\centering
\scalebox{0.9}{
\begin{minipage}{0.9\columnwidth}
\KwData{A relationship $\langle t_i, a_i, t_j, p\rangle$; and a collection of tables 
${D}_{a_i}$, each containing at least one column whose conceptual attribute is $a_i$.}
\KwResult{The inferred cardinality $Card(t_i,t_j)$  .}

${{D}}_{a_i}^{(k)} \gets \texttt{SampleTables}({{D}}_{a_i}, k)$\; \label{line:sampletables}
\ForEach{$d \in {D}_{a_i}^{(k)}$}{
    $\texttt{stats}(d, a_i) \gets \{\,
    \rho_{\text{null}}(c_{a_i}^d),\;
    \rho_{\text{multi}}(c_{a_i}^d),\;
    \rho_{\text{uniq}}(c_{a_i}^d)
    \,\}$\; \label{line:stats}
}

${D}_{a_i}^{(\text{sample})} \gets \texttt{RowSample}({D}_{a_i}^{(k)}, a_i)$\;
 \label{line:samplerows}

$Card(t_i,t_j) \gets \texttt{LLMInferCardinality}(t_i, t_j, a_i,\texttt{predicate},\\\texttt{stats}(d, a_i), {D}_{a_i}^{(\text{sample})}
)$\; \label{line:llminfer}
\Return{{$Card(t_i,t_j)$ }.}
\end{minipage}
}
\end{algorithm}

\methodG begins by sampling tables that contain columns associated with the conceptual attribute $a_i$, for computing the statistical evidence.
For the associated columns of $a_i$ in each sampled table, we compute the following summary statistics: (1)$\rho_{\text{null}}$: the proportion of missing entries; (2)$\rho_{\text{multi}}$: the proportion of cells that contain multiple entities listed within a single cell; and (3)$\rho_{\text{uniq}}$: the ratio of distinct cell values to the total number of cells in a column.

These indicators jointly characterize the completeness, variability, and multiplicity of the relationship attribute: a low $\rho_{\text{uniq}}$ and high $\rho_{\text{multi}}$ typically indicate a one-to-many relationship, whereas high $\rho_{\text{uniq}}$ and low $\rho_{\text{null}}$ suggest a one-to-one mapping. They serve as quantitative evidence guiding the subsequent cardinality inference.

To complement these statistics, \methodG further extracts 5 sampled rows from each sampled table.
Neighboring columns and their values are retained to reflect the full table context of the relationship.
In the end, a prompt is constructed with all the above information, as shown in Figure~\ref{fig:rcd}, and provided to the LLM for inference.
The prompt includes (i) a brief definition of relationship cardinality, (ii) the involved types $(t_i, t_j)$ and their predicate, and (iii) several sampled table instances, each including the summary statistics of the column associated with the attribute $a_i$ and sampled rows.
The LLM is instructed to infer the relationship cardinality from the four options -- one-to-one (1:1), one-to-many (1:n), many-to-one (n:1), and many-to-many (m:n).

\section{Schema Inference with Table Embeddings}
\label{sec:emsi}

\begin{figure*}[tb]
  \centering
  \begin{minipage}[b]{0.53\textwidth} 
    \centering
    \includegraphics[width=\linewidth]{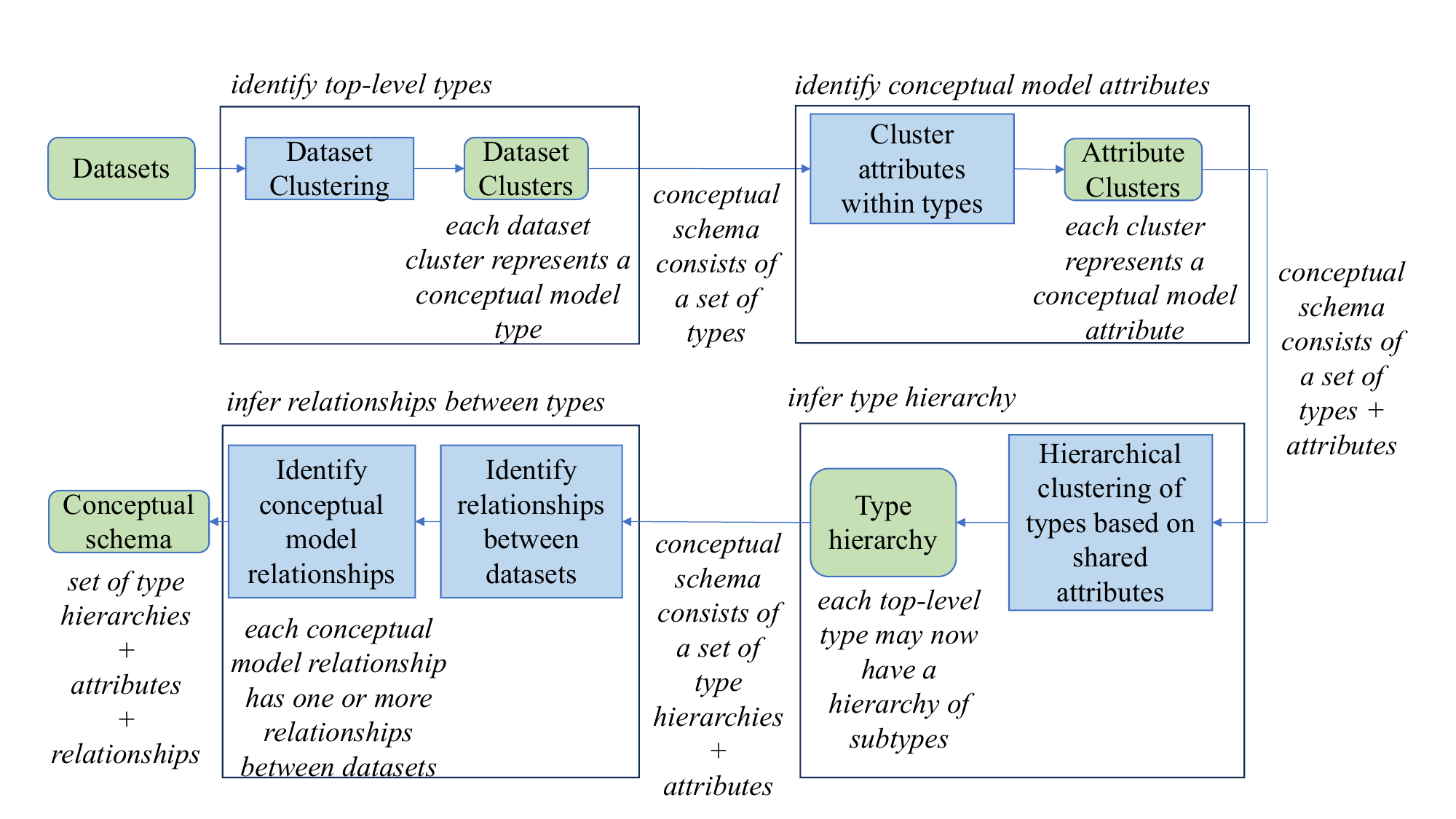}
    \caption{Summarization of the four steps of \methodE}
    \label{fig:steps}
 \end{minipage}
 \begin{minipage}[b]{0.45\textwidth}
  \centering
    \includegraphics[width=\linewidth]{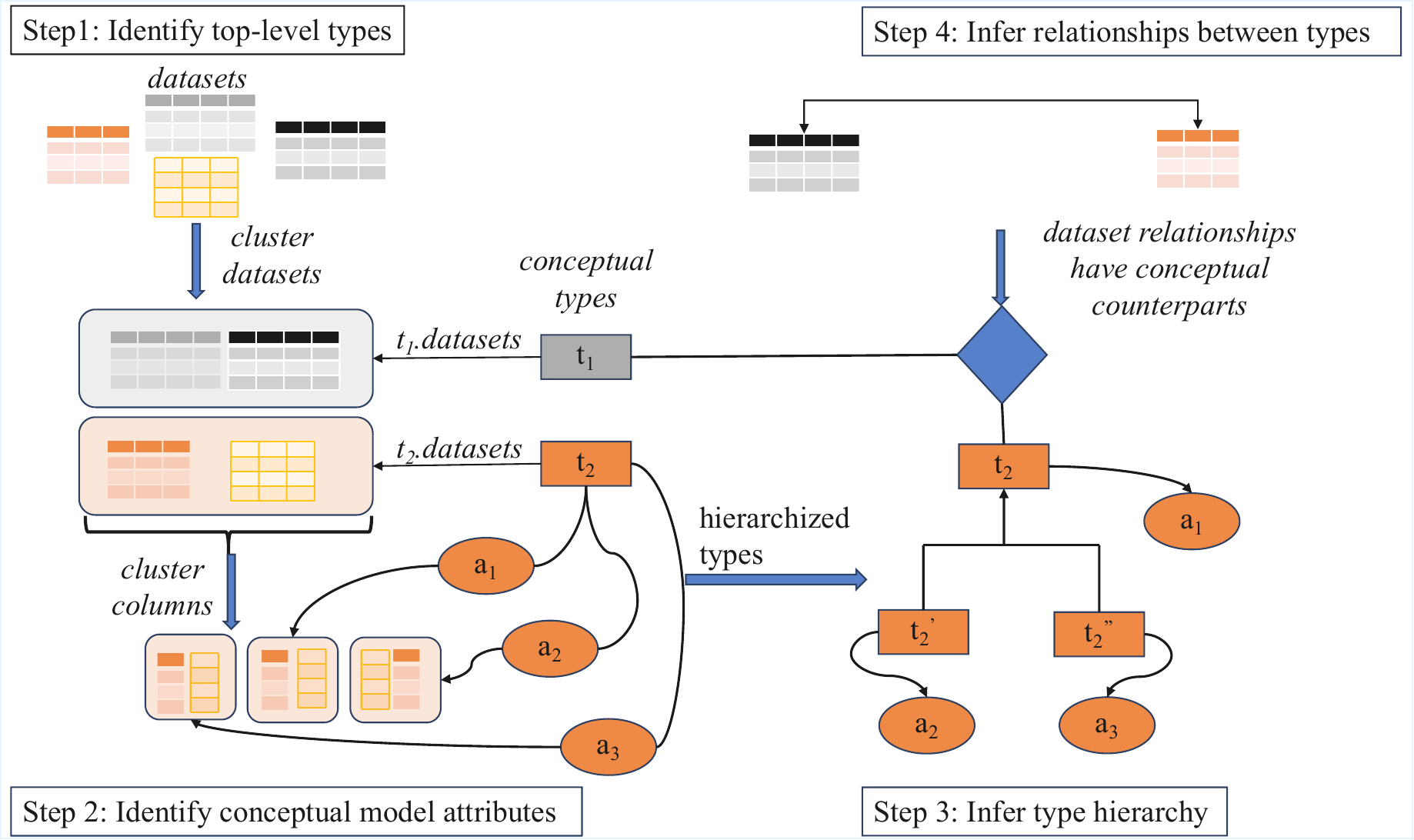}
    \caption{Demonstration of \methodE with the input and output of its four steps}
    \label{fig:step-results}
    \end{minipage}
\end{figure*}

{Repository-level schema inference with embedding models raises several challenges. 
First, different schema elements rely on different evidence, so relevant information from tables and columns must be selected and encoded. 
Second, embedded tables and columns need to be grouped into conceptual types and attributes, even though columns with similar names or values may play different roles. 
Third, once types and attributes are identified, hierarchies and relationships need to be inferred from attribute evidence, including attributes that suggest type specialisation and attribute values that indicate type-level relationships despite inconsistent instance representations.
}

{These challenges motivate the similarity-based design of \methodE. 
\methodE first represents table and column evidence in embedding spaces, and then uses similarity-driven clustering and matching to derive top-level conceptual types, conceptual attributes, type hierarchies, and relationships. 
As shown in Figures~\ref{fig:steps} and \ref{fig:step-results}, \methodE proceeds through four steps:}

\begin{itemize}[leftmargin=*]
    \item Step 1: {\it Identify top-level types}. Given a set of tables $D$, this step clusters tables in such a way that each cluster represents tables with the same top-level conceptual type $t \in T_{\text{top}}$.
    \item Step 2: {\it Identify conceptual attributes}. Given a conceptual type $t$ from Step (1) that is associated with a cluster of tables $t.datasets$, cluster the columns of these tables in such a way that each cluster of columns represents a conceptual type attribute $a \in \texttt{Atts(t)}$.
    \item Step 3: {\it Infer type hierarchy}. Given a conceptual type $t$ from Step (1) that is associated with a cluster of tables $t.datasets$, divide these tables into groups, each of which represents a direct or indirect subtype of the top-level type $t$ in the conceptual model, based on the sharing of attributes identified at Step (2).  
    \item Step 4: {\it Identify relationships between types}. Given a pair of types, either top-level types from Step (1) or the subtypes from Step (3), represented as $t_i, t_j \in T$, identify relationships between the subject columns in tables in $t_i.datasets$ and other columns of the tables in $t_j.datasets$\footnote{A {\it subject column} is a column that denotes the concept that a table is about~\cite{DBLP:journals/pvldb/VenetisHMPSWMW11}. For example, this could be the {\it title} of a {\it Event} or the {\it name} of a {\it book}.}.  When a threshold fraction of the attributes from $t_j.datasets$ are similar to some subject attribute from $t_i.datasets$, a relationship is created between the corresponding conceptual attributes in $t_i$ and $t_j$.
\end{itemize}
The rest of this section details each of these steps.

\subsection{Identify Top-level Types}
\label{sec:infer-types}

In this step, a set of clusters $C$ is created such that each $c \in C$ consists of a set of tables from $D$. Each cluster is intended to group tables that are associated with a single top-level type $t \in T_{\text{top}}$ in the conceptual schema. 
These top-level types, such as {\it Person} and {\it CreativeWork}, may subsequently be associated with subtypes, as described in Section \ref{sec:infer-hierarchies}. 
The approach involves two main sub-steps: (i) computing a \textit{subjectness score} for each column in a table using an existing method from $\text{TableMiner}^{+}$~\cite{DBLP:journals/semweb/Zhang17}, and deriving a column weight based on this score; and (ii) clustering the tables based on the similarity between their table embeddings, where each table's embedding is obtained as the weighted aggregation of the embeddings of its columns. 

Clustering of underlying tables or data items has been widely used for type identification in schema inference (e.g.,~\cite{DBLP:journals/tlsdkcs/ChristodoulouPF15,Kellou-Menouer-22,DBLP:conf/edbt/BonifatiDM22}).
Such clustering approaches typically define similarity based on features of the underlying tables, such as the Jaccard similarity of their attributes or relationships.
In \methodE, we instead cluster tables using the similarity of their {weighted table embeddings}, where each column is assigned a weight proportional to its contribution as a potential subject of the table.
This design aims to group together tables that represent the same real-world concept when inferring top-level types, while reducing the influence of differences in their descriptive columns.
For instance, we seek to infer a top-level type \textit{Person} that encompasses tables describing different perspectives, such as \textit{Actors} or \textit{Students}, even though these may have distinct attribute sets.

When clustering heterogeneous tables, \methodE 
calculates a subjectness score for each column, quantifying how strongly the column represents the main entity type of the table, and uses these scores to derive column weights when constructing the weighted table embedding.
This weighting scheme allows the model to handle tables that contain multiple entities or have no clear subject column. In such cases, columns that describe core entities receive higher subjectness weights, while descriptive columns have less impact.
As a result, table similarity is driven primarily by their entity-defining columns rather than by variations in their descriptive columns.




%
\begin{algorithm}[tb]
\caption{Identify top-level types in \methodE (Step 1)}
\label{alg:types}
\KwData{Set of tables {$D$}, {Column embedding} model M}
\KwResult{Set of table clusters}
\For {{$d\in D$}}{
   { d.colweights = SubjectnessScore(d)\; }
}
Initialize distance matrix {$\mathbf{D}$ to max for all positions\;
\For {$d_i\in D$}{
    \For {$d_j\in D$}{
        $\mathbf{D}[d_i,d_j] = sim( d_i.colweights, d_j.colweights,M)$\;     
    }
}
\Return{cluster($\mathbf{D}$)}\;}

\end{algorithm}

Given a set of tables and an embedding model {\it M}, Algorithm \ref{alg:types} returns a set of clusters, where each cluster contains a set of tables that are considered to belong to the same type, where $sim$ returns the similarity of its first two parameters using {\it M}. In the experiments, we use Agglomerative Clustering, which was selected due to its strong performance in terms of efficacy and efficiency.

\subsection{Identify Conceptual Attributes} 
\label{sec:infer-attributes}

A conceptual type is associated with many tables, which may have recurring attributes with the same semantics.  For example, $movie_1.name$ and $movie_2.title$ capture the same characteristic of a movie, and thus can be represented by a single attribute of the type \textit{Movie}. For each {top-level} type $t \in T_{\text{top}}$, the approach of Step (2): (i) accesses all the columns in the tables from which $t$ is derived, $\texttt{Atts}(t) = \{ a | d \leftarrow t.datasets, a \leftarrow d.columns\}$; (ii) clusters these columns based on their semantic similarity, where similarity is computed using an LLM encoder. A new attribute for the type $t$ is created for each column cluster. The approach is implemented as in Algorithm \ref{alg:attributes}. 

\begin{algorithm}[tb]
\caption{Identify conceptual attributes in \methodE (Step 2).}
\label{alg:attributes}
\KwData{Top-level type $t \in T_{\text{top}}$, {Column embedding model $M$}}
\KwResult{Set of column clusters}
$columns = \texttt{Atts}(t)$\;
Initialize distance matrix {$\mathbf{D}$} to max for all positions\;
\For {$c_i\in columns$}{
    \For {$c_j\in columns$}{
        ${\mathbf{D}}[c_i,c_j] = sim(c_i,c_j,M)$\;     
    }
}
\Return{cluster({$\mathbf{D}$})}\;
\end{algorithm}

In addition, we identify the \textit{subject attribute} of each type, which represents the entity instances of that type.
Specifically, we identify an attribute as the subject attribute of a type when all its member columns are recognized as named entity columns (whose cell values represent entities) and the corresponding tables assign them a high subjectness score. We set the threshold at 50\% of the total subjectness weight of the table.
If no attribute in a type satisfies this condition, the type is treated as having no subject attribute. These subject attributes serve as anchors for identifying instance-level relationships between types in the subsequent Step (4).

\subsection{Infer Type Hierarchy} 
\label{sec:infer-hierarchies}
\begin{figure}[tb]
\centering
  \centering
  \includegraphics[width=0.9\linewidth]{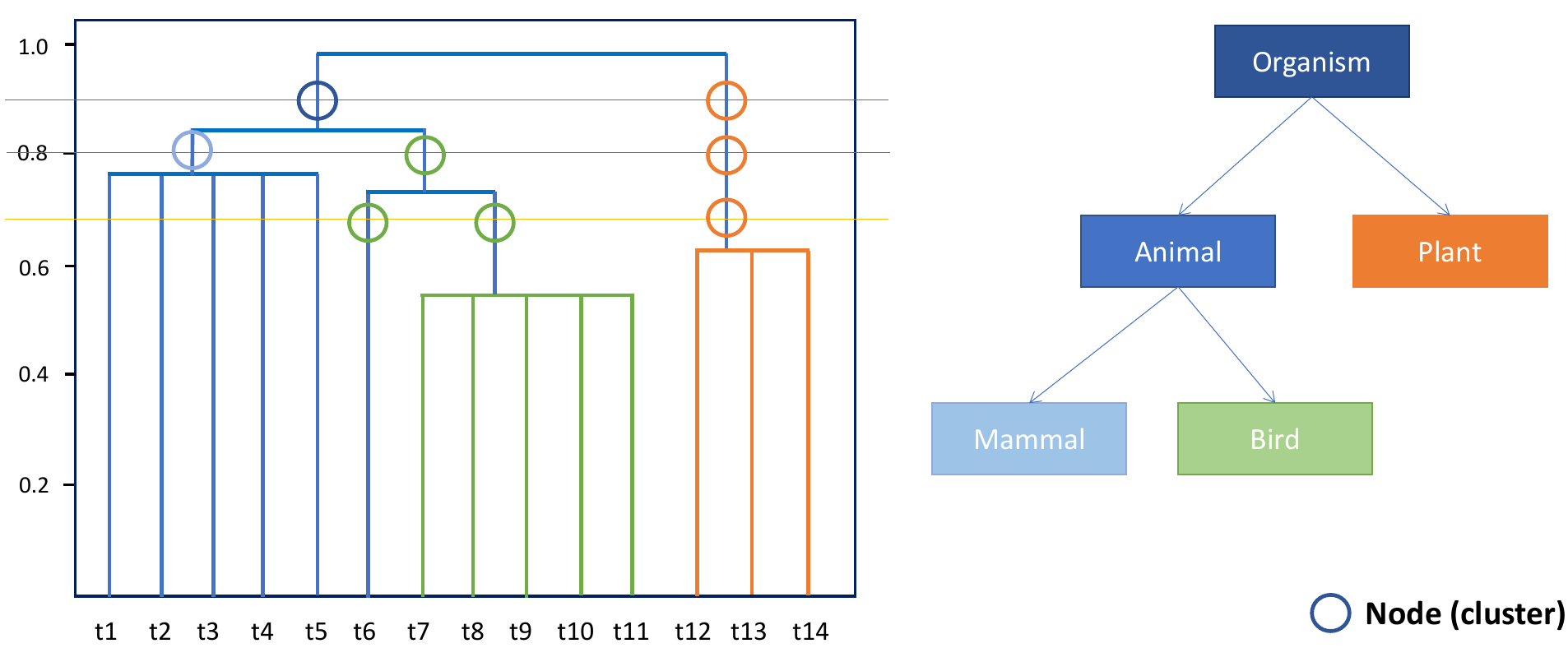}
\caption{A dendrogram example with the top-level type \textit{Organism} in WDC}
\label{fig:inheritance-example}
\end{figure}
The top-level types identified in Section \ref{sec:infer-types} are disjoint.  As they result from clustering on {weighted table embeddings}, some of the inferred types may be expected to be associated with tables that represent different perspectives. For example, a single cluster representing  \textit {CreativeWork} may be expected to include specific types of creative work, such as \textit{Movies} and \textit{Books}.  If these differences are reflected in different attributes, a type hierarchy can be created where attributes of supertypes are inherited by subtypes. For each top-level type $t \in T_{\text{top}}$, the approach includes the following steps:
\begin{itemize}
    \item Running a hierarchical clustering method on $t.datasets$, where the distance of two tables $d_i$, $d_j$ is the Jaccard similarity of their sets of conceptual attributes $d_i.conceptualAttributes$ and $d_j.conceptualAttributes$. 
    \item From the possible ways of clustering the data, represented by the dendrogram produced by hierarchical clustering, selecting those with the highest silhouette score. This has the effect of identifying groups of attributes that co-occur in several of the tables in $C$, and retaining those groups that form good clusters according to the silhouette score\footnote{The {\it silhouette score} is a technique for measuring cluster quality where there is no access to a Ground Truth, in terms of cluster coherence and separation.}.  Each of the retained groups becomes a new $Type$ in the conceptual schema, and the hierarchy of the dendrogram is reflected in the $Hierarchy$ of the conceptual model. 
\end{itemize}

\begin{algorithm}[htb]
\caption{Infer type hierarchy in \methodE (Step 3).}
\label{alg:hierarchy}
\KwData{{Top-level} type $t \in T_{\text{top}}$}
\KwResult{Type hierarchy}
Initialize distance matrix {$\mathbf{D}$} to max for all positions\;
\For {$d_i\in t.datasets$}{
    \For {$d_j\in t.datasets$}{
        $\mathbf{D}[d_i,d_j] = Jaccard(d_i.conceptualAttributes)$, \\ \quad\quad\quad\quad\quad$d_j.conceptualAttributes)$\;    
    }
}
$dendrogram = HierarchicalCluster(\mathbf{D})$\;
$maxSilhouette = maxSilhouetteScore(dendrogram)$\;
\Return $pruneDendrogram(dendrogram, maxSilhouette)$\;
\end{algorithm}
The implementation of the approach is given in Algorithm \ref{alg:hierarchy}, where $maxSilhouetteScore$ returns the highest silhouette score that can be obtained by slicing the dendrogram, $pruneDendrogram$ (Algorithm~\ref{alg:slice}) returns the hierarchy that results from pruning the dendrogram to retain only the clustering results that have a silhouette score greater than $(maxSilhouette - \Delta)$, and $\Delta$ helps determine the minimum silhouette score for the resulting clusters when slicing the dendrogram. This ensures that we do not retain layers of the hierarchy with clusters of poor quality. 
Figure \ref{fig:inheritance-example} illustrates the hierarchy inference process in practice for the top-level type \textit{Organism} in WDC. The dendrogram results from the clustering of 14 tables, $d_1$ to {$d_{14}$}, and three slices are identified with a silhouette score within $\Delta$ of the highest silhouette score. Each slice is colored in the same way as the corresponding types in the ground truth hierarchy. 
The slicing algorithm in Algorithm~\ref{alg:slice} starts from the top of the dendrogram and makes its way downward. Upon slicing, if the resulting clusters have average silhouette score within the predefined interval, they are included in the resulting \textit{hierarchy}. 

\begin{algorithm}[htb]
\caption{Prune the dendrogram.}
\label{alg:slice}
\KwData{$dendrogram$; $maxSilhouette$;}
\KwResult{Type hierarchy}
Initialize $hierarchy$\;
$current_y$ = max(y values in $dendrogram$);\\
\While{$current_y >=0$}{

$clusters$ = getClustersAtHeight($dendrogram$, $current_y$);\\
\If{$getSilhouette(clusters)>(maxSilhouette-\Delta)$}{
 \For {$cluster$ in $clusters$}{
  \If {$cluster$ is not a single table}{
  $tables$ = getTable($cluster$);\\
  $parentCluster$ = findParentCluster($hierarchy$, $cluster$);\\
  \If {$parentCluster \; exists$ }{
   AddEdge($hierarchy$, $parentCluster$, $cluster$);\\}
   \Else{addRoot($hierarchy$, $cluster$)}
  }
 }
}
 $current_y -= \delta$;\\
}
\Return{$hierarchy$}\;
\end{algorithm}


\subsection{Identify Relationships} 
\label{sec:infer-relationships}

We now describe how relationships can be derived between the types identified in Sections \ref{sec:infer-types} and \ref{sec:infer-hierarchies}.  For example, if we have identified conceptual types {\it Person} and {\it Movie}, we might expect to infer a relationship that links a movie with its \textit{Cast} or \textit{Director}.  The approach is presented in Algorithm \ref{alg:relationships} with details.
\begin{algorithm}[htb]
\caption{Identify relationships in \methodE (Step 4).}
\label{alg:relationships}
\KwData{Set of conceptual types $T$, column embedding model $M$, similarity threshold $\delta_1$, fraction threshold of non-subject conceptual attribute $\delta_2$ }
\KwResult{Relationships}
$relationships = \{\}$ \;
\For {$t_i \in T$} { 
    $csa_i.columns = Columns(t_i.csa_i)$\; 
    \For {$t_j \in T$} { 
    \If{$t_i \neq t_j$}{
       \For{$ca_j \in \texttt{Atts}(t_j)$}{
         $ca_j.columns = Columns(t_j.ca_j)$\; 
         $relatedColumns =\{\}$\; 
         \For {$c_i\in csa_i.columns$}{ 
            \For {$c_j\in ca_j.columns$}{ 
               \If {($sim(c_i,c_j,M) > \delta_1 \& c_j \not \in relatedColumns$)} {
                Add $c_j$ in $relatedColumns$\;
            }
             }
         }
         \If{$ \frac{|relatedColumns|}{|ca_j.columns|}> \delta_2$}{
                         $relationships = relationships \cup \{\langle csa_i, ca_j\rangle$\}\;
         }
      }
    }
  }
}
\Return{relationships}\;
\end{algorithm}
Briefly, for a subject conceptual attribute $csa_i$ of a type $t_i$ and a non-subject conceptual attribute $ca_j$ of another type $t_j$, a relationship is established if the two attributes' associated table columns, i.e., $csa_i.columns$ and 
$ca_j.columns$, are close. This means a threshold fraction of the columns in $ca_j.columns$ are found to be similar to the columns in $csa_i.columns$, based on a column embedding model $M$.



There are two thresholds that have been set empirically: $\delta_1$ indicates 
how similar a pair of columns should be to be considered as evidence of similarity among the conceptual model attributes; and $\delta_2$ is the fraction of the columns that need to be similar to provide confidence in the creation of a conceptual level relationship. This supports the notion that a conceptual model relationship should occur frequently in the associated table columns.


 \section{Experiments}
\label{sec:exp}
\subsection{Experiment setup}
We implemented both \methodG and \methodE in Python (v3.13) using PyTorch and the Hugging Face Transformers library. All experiments are conducted on a single NVIDIA A100 GPU (80 GB) with dual AMD EPYC CPUs and 512 GB RAM. To account for randomness, each experiment is run five times and we report the mean.

\textbf{\methodG settings.}
For \methodG, we employ four representative LLMs for evaluation: Llama-3.1-8B, Qwen2.5-14B, GPT-3.5-turbo (GPT-3.5 for short), and GPT-4.
We select models that span a broad range of capacities and accessibility: from smaller open-weight models (Llama 3.1 and Qwen2.5) to larger proprietary ones (GPT-3.5 and GPT-4) to assess the scalability and generalization of \methodG across open and closed ecosystems.
Each model is invoked statelessly with a generation temperature of 0.1. 
Llama-3.1-8B and Qwen2.5-14B run locally on our hardware setup, while GPT-3.5/ 4 are accessed via the OpenAI API. 

{
\methodG involves three step-specific parameters. 
In {Step 1}
, we set the maximum depth of the inferred type path to $N=5$, one level deeper than the ground-truth hierarchy (which has a depth of at most 4), allowing sufficient slack for capturing finer-grained intermediate types. 
In {Step 2}
, an attribute is promoted to its parent type $t$ if it appears in at least a proportion $\theta = 0.9$ of $t$'s direct child types. 
In {Step 3}
, the LLM selects the top-$K = 3$ candidate top-level types $T'_{\text{top}}$ from the type hierarchy for subsequent relationship inference.
}


\textbf{\methodE settings.}
\methodE is agnostic to column embedding models.
We employ three representative fine-tuning-based models: Starmie \cite{DBLP:journals/pvldb/FanWLZM23}, DeepJoin \cite{DBLP:journals/pvldb/Dong0NEO23}, and SwAV \cite{wu2025taxonomyinferencet}, following the  parameter settings in \cite{wu2025taxonomyinferencet}. 
They cover different learning paradigms: Starmie uses contrastive learning for column-level similarity \cite{DBLP:conf/icml/ChenK0H20}, DeepJoin focuses on joinability-based supervision, and SwAV adopts clustering-based self-supervision \cite{DBLP:conf/nips/CaronMMGBJ20}. 
For reference, we additionally include SBERT~\cite{DBLP:conf/emnlp/ReimersG19} and Unicorn~\cite{DBLP:journals/sigmod/FanTLWDJGT24}, two pretrained Transformer-based encoders trained on general-purpose and schema-matching tasks, respectively, without further fine-tuning. 

{

{For \methodE, we use WDC as a development dataset for parameter selection to avoid tuning separately on each benchmark. 
In Step~3, $\Delta$ is selected from $[0.05,0.50]$ with increments of $0.05$ based on PTCS (cf. Section~\ref{sec:PTCS}) and the resulting hierarchy depth, giving $\Delta=0.15$, which is fixed for all datasets and embedding variants. 
In Step~4, because similarity scores are not directly comparable across embedding models, we select model-specific thresholds $(\delta_1,\delta_2)$ on WDC by maximizing F1-score subject to Precision $>0.5$ and Recall $>0.05$, and then fix them for the remaining datasets. 
The selected thresholds are: SBERT $(0.75,0.15)$; DeepJoin $(0.75,0.20)$; Starmie $(0.45,0)$; and SwAV $(0.30,0)$.}



}

\subsection{Benchmarks}
\label{sec:benchmark}

\begin{table*}[tb]
\centering
\resizebox{0.9\textwidth}{!}{
\begin{tabular}{|c|c|c|c|c|c|c|c|}
\hline
Benchmark & \#Tables & \#Columns & \#Top-level Types& \#Types& \#Layer &\#Attribute&\#Relationships\\ \hline 
WDC       & 602      & 4200      & 7                  & 43       & 4       &      475&46\\ \hline
GDS       & 660      & 15195     & 6                  & 53          & 4       &   1187&59\\ \hline
OpenData (Small)  & 158      &    3799  & 4                & 24         & 2  &  340   &7\\ \hline
OpenData (Large)   & 10361     &   354666   & 6                  & 53          & 4    &   -  & -\\ \hline
\end{tabular}
}
\caption{Benchmark information}
\label{tab:benchmark}
\end{table*}

Table~\ref{tab:benchmark} summarizes the statistics of the benchmarks used in our experiments. 
They collectively cover heterogeneous table sources and different scales of both tables and conceptual schemas, enabling a comprehensive evaluation of both \methodG and \methodE as well as the types, type hierarchy, conceptual attributes, and relationships of the inferred schema.

\begin{itemize}
\item \textit{WDC benchmark.} 
Comprising 602 tables drawn from Web Data Commons~\cite{springer/ISWC14/WDC} and T2DV2~\cite{DBLP:conf/EDBT17/T2DV2}, this benchmark provides type labels from the original datasets, which are aligned with types in Schema.org, and attribute labels that are manually added following the Web Data Commons annotations. 
The corpus exhibits a long-tailed distribution across 43 conceptual types, of which 21 types are associated with fewer than 5 table instances.

\item \textit{GDS benchmark.}
This benchmark contains 660 web tables from Google Dataset Search~\cite{conf/WWW/DanBrickey19}, {where the type labels were originally derived from Schema.org and further refined manually, attributes and inter-type relationships were manually annotated.
Each top-level type includes at least 100 table instances, and each lowest level type has a minimum of 10 table instances.}

\item \textit{OpenData (Small) (OD Small).} 
{This benchmark contains 158 tables sampled from open data portals~\cite{ukopen,usopen,ausopen}. A key challenge in this dataset is that many tables are "multi-subject tables": they lack a single, dominant subject column, even though the table as a whole represents a single latent entity type (e.g., a \textit{Transaction} table). We use this dataset to evaluate how well \methodG and \methodE infer these latent entity types and their hierarchy from tables.
Each table is annotated with the most specific Schema.org type that serves as a common type for all entities in the table. The full type path of each table is then derived by traversing the Schema.org hierarchy upward to the root type “\textit{Thing}”. We merge these paths to form ground truth type hierarchy used for evaluation. Attributes and inter-type relationships are also manually annotated.
}

\item \textit{OpenData (Large) (OD Large).}
This benchmark contains a large-scale collection 10,361 tables from the same open data portals as OD Small. 
It follows the same annotation strategy as OD Small, including the annotations of Schema.org types and construction of type hierarchy.
To assess scalability and generalization, this benchmark is used exclusively for type and type hierarchy inference. 
Attributes and inter-type relationships are not annotated, as exhaustive manual labeling at this scale would be prohibitively expensive.

\end{itemize}


\subsection{Baselines}
Both \methodG and \methodE aim to infer conceptual schemas but follow different inference paradigms. 
As their procedures differ, we do not compare them with step-by-step alignment. 
Instead, we compare the resulting schema components of \methodG and \methodE: (i) types and the type hierarchy, (ii) conceptual attributes, and  (iii) relationships and their cardinalities, across different LLM settings.
Considering conceptual schema inference is a new problem we recently proposed, there is a shortage of {overall} comparable baselines. {For type and type hierarchy inference, we use GeTT~\cite{wu2025taxonomyinferencet} as a baseline.
GeTT is a generative LLM-based approach that first predicts entity types for individual tables and then constructs the is-a relationships among them using strategies such as breadth-first search.
For conceptual attribute inference, we adopt SI-LLM~\cite{wu2025schema} as a baseline. This method integrates LLM-based attribute naming with LLM-guided grouping for name resolution.
Specifically, it first generates an attribute name for each table column, de-duplicates the resulting names, and then uses the LLM to group similar names together. 
Each group is assigned a unified attribute name.
}

\subsection{Evaluation Metrics}

\noindent\textbf{Metrics for the top-level types and conceptual attributes.} 
Both top-level type inference and conceptual attribute inference can be viewed as clustering tasks, and thus
we adopt two standard metrics: the Rand Index (RI) and Purity.
RI measures the pairwise agreement between the predicted and the {Ground-Truth (GT)} clusters:
$$
Rand\:Index = \frac{TP + TN}{TP + FP + FN + TN},
$$
{where TP (true positive) denotes pairs that are in the same cluster in both prediction and GT; FP (false positives) denotes pairs that are clustered together in the prediction but not in the GT; FN (false negatives) denotes pairs that are in different clusters in the prediction but in the same cluster in the GT; TN (true negatives) denotes pairs that are in different clusters in both prediction and GT. }

Purity evaluates the homogeneity of each predicted cluster with respect to the ground-truth labels:
$$
Purity = \frac{1}{K} \sum_{i=1}^{K} 
\frac{\max_j |C_i \cap L_j|}{|C_i|},
$$
{where \(C_i\) denotes the \(i\)-th predicted cluster, \(L_j\) denotes the set of instances belonging to the \(j\)-th GT label, 
\(|C_i|\) is the size of cluster \(C_i\), and \(\max_j |C_i \cap L_j|\) is the number of instances in \(C_i\) assigned to the most frequent GT label.}
{When evaluating {top-level type inference}, the instances are {tables} and the GT type labels correspond to {top-level types}. 
When evaluating {attribute inference}, the instances are {columns} and the labels correspond to {GT attributes}.}
\label{eqn:Purity}



\noindent\textbf{Metrics for the type hierarchies.} To assess the structural richness of the inferred hierarchy, we report two basic statistics: the total number of inferred types (\textit{T\#}) and the maximum depth of the hierarchy from \textit{Thing} to the leaf types (\textit{L\#}).
A larger number of types or greater depth indicates a more fine-grained hierarchy with richer semantics.

Besides, we further evaluate how well the inferred hierarchy aligns with the GT. 
Rather than evaluating the correctness of every is-a edge, we assess the overall consistency between the generated type hierarchy and the type hierarchy of the GT schema.
This is also because there is often more than one type hierarchy that can model the conceptual semantics of a given set of tables.
{For example, given the type \textit{Livestream Concert} under top-level type \textit{Event}, one possible hierarchy may organize subtypes by topic, such as \textit{Music Event} → \textit{Livestream Concert},
while another may organize them by format, such as \textit{Online Event} → \textit{Livestream Concert}.
Both hierarchies describe the same set of Livestream Concert related tables but differ in their intermediate types and structures.}
To this end, we propose the {Path-based Tree Consistency Score (PTCS)}\label{sec:PTCS}. 
Given a complete path $p = [t_1, t_2, \dots, t_n]$ in the inferred hierarchy $H$, where $t_1$ is a top-level type and $t_n$ is a leaf type, 
each of its types $t_i$ is aligned to a ground-truth type $t_i'$ in the GT hierarchy $H'$.
{Specifically, $t_i'$ is determined as the {most frequent ground-truth type among the tables associated with $t_i$ in $H$}.}
This yields a corresponding sequence $p' = [t_1', t_2', \dots, t_n']$. 
We then compute PTCS of this path $p$:
$\text{PTCS}(p) = \frac{L(p')}{|p'|},$
where $L(p')$ denotes the length of the longest valid ancestor–descendant subsequence of $p'$ in $H'$. 
The overall PTCS for $H$ is obtained by averaging the PTCSes across all its complete paths. 
A higher PTCS indicates greater consistency between the inferred and the GT hierarchies.

\noindent\textbf{Metrics for the relationships and their cardinalities.} 
We evaluate the inferred inter-type relationships by comparing them with the GT set, and calculate three standard metrics: Precision (\(P\)) as the proportion of the inferred relationships that are among the GT set, Recall (\(P\)) as the proportion of the GT relationships that are among the inferred relationships, and the F1 score as the harmonic mean of Precision and Recall. 


For relationship {cardinality}, we consider those correctly identified relationships, and calculate Accuracy which is the proportion of relationships whose predicted cardinality labels (one-to-one, one-to-many, many-to-one, or many-to-many) match the GT.

\subsection{Results on Types and Type Hierarchy}

We report and analyze results in two dimensions: the top-level types, and the overall consistency of the type hierarchy w.r.t. the GT.

\begin{table*}[tb]
\centering
\resizebox{0.75\textwidth}{!}{
\begin{tabular}{c|cc|cc|cc|cc}
\hline
\multicolumn{1}{l|}{}       & \multicolumn{2}{c|}{{WDC}} & \multicolumn{2}{c|}{GDS}                               & \multicolumn{2}{c|}{OD Small}                             & \multicolumn{2}{c}{OD Large}                              \\ \cline{2-9} 
\multicolumn{1}{l|}{}       & {RI}   & {Purity}   & {RI}               & {Purity}            & {RI}                     & {Purity}                 & {RI}                     & {Purity}                 \\ \hline
\methodG (Llama-3.1-8B)        & 0.812         & 0.965             & 0.672     &0.926                      & 0.942                           & 0.963                           & 0.683 & {0.937}  \\
\methodG (Qwen2.5-14B)        & \textbf{0.841}         & 0.979             & 0.69 & 0.943 & \underline{0.967}  & 0.968          & \underline{0.786}  & \underline{0.944}        \\
\methodG (GPT-3.5)       & 0.767         &\underline{0.984}             & \textbf{0.816} & \textbf{0.978}& \textbf{0.99}                   & \textbf{0.993}                  & \textbf{0.834}                  & \textbf{0.964}                  \\
\methodG (GPT-4)               & \underline{0.824}         & \textbf{0.989}             & \underline{0.715}& \underline{0.968} & 0.958                           & \underline{0.978}  & 0.746                           & 0.941                         \\ \hline
{\methodE (SBERT)}       & 0.795         & 0.823             & 0.803                     & 0.689                      & \underline{0.946}  & \underline{0.965}  & \textbf{0.893}                  & \textbf{0.895}                  \\
{\methodE (Starmie)}     & 0.783         & 0.776             & 0.819                     & 0.732                      & 0.897                           & 0.873                           & 0.854                           & 0.793                           \\
{\methodE (DeepJoin)}    & \underline{0.862}         & 0.852             & \textbf{0.905}                     & \underline{0.873}                      & 0.926                           & 0.962                           & 0.883                           & 0.88                            \\
{\methodE (Unicorn)}     & 0.734         & \underline{0.89}              & 0.709                     & \textbf{0.903}                      & 0.756                           & 0.917                           & -                               & -                               \\
{\methodE (SwAV)}        & \textbf{0.887}         & \textbf{0.891}             & \underline{0.894}                     & 0.864                      & \textbf{0.952}                  & \textbf{0.977}                  & \underline{0.892}  & \underline{0.874}  \\ \hline
{GeTT (Qwen2.5-14B)} & 0.727         & 0.738             & 0.784                     & 0.61                       & 0.796                           & 0.732                           & 0.836                           & 0.797                           \\
{GeTT (Qwen2.5-32B)} & 0.814         & 0.93              & 0.751                     & 0.673                      & \textbf{0.814}                  & 0.728                           & 0.809                           & 0.782      \\
{GeTT (GPT-3.5)}     & 0.516         & 0.747             & \textbf{0.852}                     & 0.526                      & 0.739                           & \underline{0.892}  & 0.879                           & \textbf{0.914}                  \\
{GeTT (GPT-4)}       & \underline{0.976}         & \textbf{0.961}             & 0.76                      & \textbf{0.9}                        & 0.716                           & 0.855                           & \textbf{0.912}                  & \underline{0.906}  \\
{GeTT (DeepSeek-R1)} & \textbf{0.98}          & \underline{0.957}             & \underline{0.819}                     & \underline{0.787}                      & \underline{0.803}  & \textbf{0.912}                  & \underline{0.907}  & 0.898                           \\
                     
\hline
\end{tabular}}
\caption{Results of the top-level types. The best result of different settings for each method is bolded, and the second best is underlined.}
\label{tab:topResults}
\end{table*}

\noindent\textbf{Top-level types.}
\label{sec:eval-types}
As top-level types depend on the granularity and categorization dimension of the hierarchy, the evaluation should therefore be regarded as relative rather than absolute.
For instance, a table about movies could be labeled broadly as \textit{CreativeWork} or more specifically as \textit{Movie}. Consequently, comparative evaluation across methods is often more informative than absolute accuracy, since reasonable predictions may correspond to either more general or more specific types.

Results on all four benchmarks are presented in Table~\ref{tab:topResults}. We observe three consistent phenomena. 
(i) \methodG achieves the highest Purity across datasets, improving over GeTT by 2.8–8.1\% and over \methodE by 1.6–9.8\%. 
(ii) \methodE yields higher Rand Index on three out of four benchmarks, while \methodG achieves the highest Rand Index only on \textit{OD Small}, with a 17\% gain over GeTT. 
(iii) \methodE performs comparably or slightly better than GeTT on {GDS} and {OD Large}. 
{Taken together, these results indicate that \methodG better distinguishes top-level types, whereas \methodE yields more accurate global clustering results.}
 
To explain \methodG's performance, we examine how it models types and their hierarchies.
The high Purity is attributed to the LLM's accuracy during per-table type path inference. The model consistently generates paths where the inferred types align with the conceptual domain of each table. This ensures that tables of the same type 
are grouped together. Consequently, the table clusters defined by top-level types are highly homogeneous. By contrast, the lower RI arises because the top-level type associated table clusters in the global hierarchy are not disjoint. This is a direct result of \methodG, which explicitly prompts the LLM to output one or multiple type paths per table (cf. Section~\ref{sec:perTableHier}). Consequently, the model assigns a single table to multiple, semantically valid, parallel paths. 
For example, a \textit{Book} table may be placed under both \textit{Product} (as it can be sold) and \textit{CreativeWork} (as an artistic creation). When these per-table paths are merged and pruned globally, the \textit{Product} and \textit{CreativeWork} clusters overlap because both include the same \textit{Book} tables. 

\methodG's performance is robust across different LLM backbones, with RI varying by 2–13\% and Purity by 2–5\%. In contrast, GeTT exhibits larger variation (RI 8–24\%, Purity 18–30\%). Furthermore, \methodG's performance does not strictly increase with model scale; for instance, GPT-4  does not yield the highest scores, while the much smaller Qwen2.5-14B often ranks second best. This suggests \methodG's reasoning strategy makes it less dependent on the specific LLM backbone choice compared to GeTT.


For \methodE, the performance of top-level type inference depends directly on the embedding model used for clustering.
SwAV and DeepJoin achieve the highest RI, surpassing Starmie, SBERT, and Unicorn by 9\%, 10\%, and 13\%, respectively. 
SwAV benefits from its contrastive clustering objective, which enforces consistency between augmented views of semantically similar columns, leading to compact and well-separated clusters. 
DeepJoin also improves clustering by fine-tuning on positive column pairs. Here, a pair of columns is considered positive if their cosine similarity within the dataset exceeds the predefined threshold 0.9.
In contrast, Starmie’s within-batch contrastive training is less effective for distinguishing attributes that share similar surface values but differ semantically. For instance, album titles (\textit{``The Mozart Album''}) versus event names (\textit{``Mozart Gala''}). 
Unicorn produces highly pure but fragmented clusters due to its sparse graph construction, which leaves many attributes unmatched. 
These findings suggest that embedding-based top-level type inference in \methodE is sensitive to how the training strategy defines semantic similarity in learning the embedding model. 



\begin{table*}[tb]
\centering
\resizebox{0.85\linewidth}{!}{
\begin{tabular}{c|ccc|ccc|ccc|ccc}
\hline
\multicolumn{1}{l|}{}       & \multicolumn{3}{c|}{{WDC}}          & \multicolumn{3}{c|}{{GDS}}          & \multicolumn{3}{c|}{{OD Small}}                                & \multicolumn{3}{c}{{OD Large}}                                                   \\ \cline{2-13} 
\multicolumn{1}{l|}{}       & {PTCS} & {T\#} & {L\#} & {PTCS} & {T\#} & {L\#} & {PTCS}                    & {T\#}                 & {L\#} & {PTCS}                    & {T\#}                   & {L\#}                 \\ \hline
\methodG (Llama-3.1-8B)        & 0.687        & \textbf{426}          & \underline{4}            & 0.682        & \textbf{393}          & \textbf{5}            & 0.782                           & \underline{52}  & \underline{2}            & 0.633                           & \textbf{1358}                  & \underline{10} \\
\methodG (Qwen2.5-14B)        & 0.759        & \underline{389}          & \textbf{5}            & 0.707        & \underline{306}          & \textbf{5}         & \textbf{0.868}                  & \textbf{58}                  & \textbf{3}   & \underline{0.706}  & \underline{1275}  & \textbf{11}                  \\
\methodG (GPT-3.5)       & \underline{0.776}        & 208          & 3            & \underline{0.771}        & 202          & \underline{4}            & 0.804                           & 47                           & \underline{2}            & 0.692                           & 842                            & 8                            \\
\methodG (GPT-4)               & \textbf{0.847}        & 260          & 3            & \textbf{0.791}        & 195          & 3            & \underline{0.843}  & 39                           & \underline{2}            & \textbf{0.721}                  & 752                            & 6                            \\ \hline
{\methodE (SBERT)}       & \textbf{1}            & 158          & \underline{5}            & \textbf{0.929}        & \underline{231}          & \underline{5}            & \textbf{0.961}                  & 27                           & \underline{3}            & \textbf{0.845}                  & \underline{936}   & \underline{7}                            \\
{\methodE (Starmie)}     & \textbf{1}            & 117          & 2            & \underline{0.895}        & 134          & 3            & \underline{0.924}  & 14                           & 2            & 0.813                           & 743                            & 5                            \\
{\methodE (DeepJoin)}    & 0.836        & \underline{162}          & 4            & 0.848        & 214          & \underline{5}            & 0.896                           & \underline{32}  & \textbf{4}   & \underline{0.821}  & 897                            & \underline{7}  \\
{\methodE (Unicorn)}     & -            & -            & -            & -            & -            & -            & -                               & -                            & -            & -                               & -                              & -                            \\
{\methodE (SwAV)}        & \underline{0.856}        & \textbf{213}          & \textbf{6}            & 0.8          & \textbf{276}          & \textbf{7}            & 0.878                           & \textbf{34}                  & \textbf{4}   & 0.783                           & \textbf{1228}                  & \textbf{10}                  \\ \hline
{GeTT (Qwen2.5-14B)} & 0.602        & 352          & 4            & \underline{0.562}        & 294          & 4            & \underline{0.677}  & \textbf{45}                  & \textbf{4}   & 0.453                           & \textbf{1344}                  & \textbf{7}                   \\
{GeTT (Qwen2.5-32B)} & 0.681        & \underline{356}          & \textbf{5}            & 0.493        & 295          & \textbf{6}            & 0.634                           & \underline{36}  & \textbf{5}   & \textbf{0.545}                           & \underline{ 1258} & 6                            \\ 
{GeTT (GPT-3.5)}     & 0.49         & 354          & 4            & 0.437        & 293          & \underline{5}           & \textbf{0.683}                  & 35                           & 4            & 0.457                           & 1034                           & 5                            \\
{GeTT (GPT-4)}       & \textbf{0.816}        & \textbf{358}          & 5            & 0.425        & \underline{296}          & \underline{5}            & 0.645                           & 28                           & 3            & 0.516                           & 1162                           & 6                            \\
{GeTT (DeepSeek-R1)} & \underline{0.813}        & 348          &\textbf{ 5 }           & \textbf{0.596}        & \textbf{302}          & \underline{5}            & 0.636                           & 32                           & 3            & \underline{0.532}                           & 1216                           & \underline{6}  \\

\hline
\end{tabular}
}
\caption{Results of the type hierarchy. The best result of different settings for each method is bolded, and the second best is underlined.} 
\label{tab:HierarchyR}
\end{table*}

\noindent \textbf{Type hierarchy.}\label{sec:eval-hierarchies}  
Table~\ref{tab:HierarchyR} compares the inferred type hierarchies produced by \methodG, \methodE and GeTT. 
\methodE achieves the highest PTCS and exhibits strong robustness, consistently maintaining a score above 0.80 across all the datasets. 
Comparing its best performance with \methodG's best, it outperforms \methodG by 9.3\% to 15.3\%.
\methodG's performance is less stable than \methodE, with scores ranging from 0.70 to 0.85 when different backbone LLMs are used. Despite this variance, \methodG still consistently surpasses GeTT, outperforming it by 3.1\% to 19.5\% when both methods are compared using the same LLM backbone.

Yet, these consistency scores mask notable structural differences. \methodG's hierarchies are consistently larger than \methodE's. \methodG generated 213, 117, 24, and 130 more distinct types on the four datasets, respectively, indicating more comprehensive branching. In terms of depth, however, \methodG's hierarchies were often shallower. Its average path depth (across backbones) was 0.5 to 1.0 levels shallower than \methodE's on the first three datasets, though it was 1.5 levels deeper on OD Large.

In summary, \methodE and \methodG exhibit a clear structural trade-off. \methodE produces deeper, narrower, and more vertically consistent hierarchies, as reflected by its superior PTCS. \methodG, in contrast, generates wider hierarchies with more comprehensive branching. This breadth, however, comes at the cost of being often shallower and demonstrating lower structural consistency (lower PTCS).

\methodG's lower PTCS stems from the richer, multi-path hierarchy it generates. 
Consider tables describing \textit{Music Event} as an example. In the GT hierarchy, these tables fall under the \textit{Music Event} type, a descendant of the \textit{Event} top-level type. The inferred paths from \methodG are: \noindent{\ttfamily\footnotesize Thing → Product → CulturalProduct → MusicEvent;  \\Thing → Event →  MusicEvent}. 
 This divergence is penalized by the PTCS matching mechanism, which matches each inferred type to the most frequent GT type among its associated tables. In this scenario, the inferred \textit{CulturalProduct} aligns to GT type \textit{Event}. Its inferred parent type, \textit{Product} (assuming it groups a majority of other \textit{Product} tables), is matched to GT type \textit{Product}. Consequently, the matched type path corresponding to \methodG's inferred structure becomes \noindent{\ttfamily\footnotesize Thing → Product → \\Event → Music Event}. This resulting path is structurally inconsistent with the GT hierarchy, where \textit{Product} and \textit{Event} are disjoint top-level types. 

$\methodE$ achieved the best overall PTCS. 
This success is attributable to its core mechanism of grouping tables via shared conceptual attributes (cf. Section~\ref{sec:eval-attributes}), which are often strong indicators of a table's type.
However, relying on shared attributes sometimes may introduce failures that can lower PTCS.
Even when attributes are informative, they may be insufficient to separate sibling types that inherently share highly similar attributes under the same top-level type.

For instance, Figure~\ref{tab:OperaHouse} and~\ref{tab:Hotel} respectively describe the types \textit{OperaHouse} and \textit{Hotel}. 
They share the attributes \textit{location}, \textit{established\_year} and \textit{rating\_score}, while \textit{Hotel} uniquely contains the attribute \textit{source}.
\methodE performs hierarchical clustering based on attribute overlap and tends to preserve layers of the dendrogram where small attribute differences, such as the presence of \textit{source}, are interpreted as evidence that one type is a more specific version of another. As a result, it may incorrectly infer a parent–child relationship by treating \textit{Hotel} as a subtype of \textit{OperaHouse}, which contradicts the GT hierarchy and thus reduces PTCS.
This is why \methodE's variant can get higher PTCS scores while its inferred hierarchy is shallow: fewer layers mean fewer chances to make wrong parent–child relationships.

\begin{figure*}[t]
\centering
\begin{minipage}[t]{0.32\textwidth}
\vspace{0pt}
\centering
  \includegraphics[width=\linewidth]{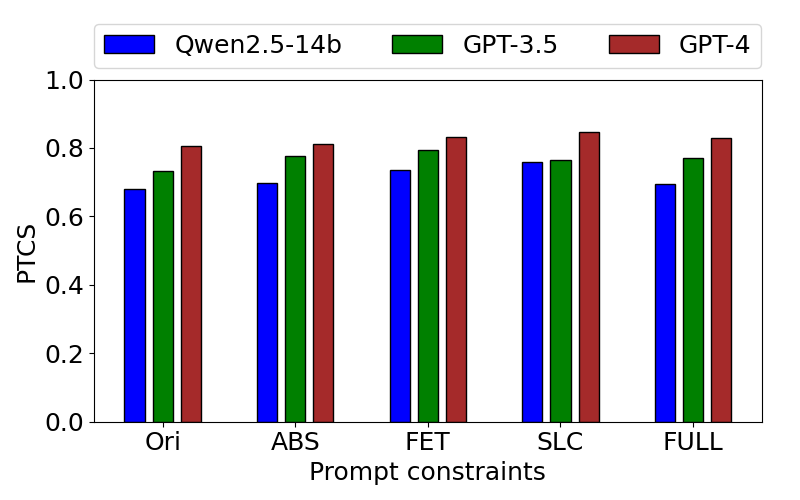}
  \vspace{-2em}
\caption{PTCS of the hierarchies inferred using different prompt constraints on WDC.}
\label{fig:fig1}
\end{minipage}
\hfill
\begin{minipage}[t]{0.32\textwidth}
\vspace{0pt}
\centering
 \includegraphics[width=\linewidth]{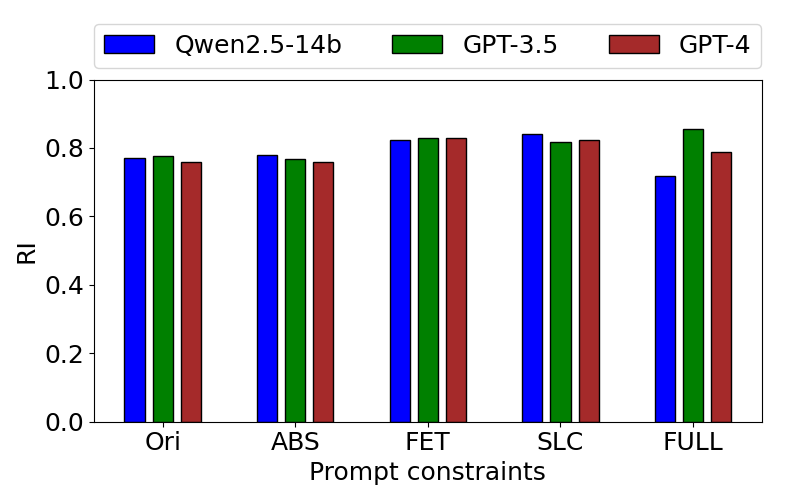}
   \vspace{-2em}
\caption{RI of the top-level types inferred using different prompt constraints on WDC.}
\label{fig:fig2}
\end{minipage}
\hfill
\begin{minipage}[t]{0.33\textwidth}
\vspace{6pt}
\centering
\resizebox{1\linewidth}{!}{
\begin{tabular}{lccc}
\toprule
Dataset & w/o & w/ & Imp. \\
\midrule
WDC      & 0.623 & 0.776 & +0.153 \\
GDS      & 0.718 & 0.771 & +0.053 \\
OD Small & 0.726 & 0.804 & +0.078 \\
OD Large & 0.557 & 0.692 & +0.135 \\
\bottomrule
\end{tabular}
}
\caption{PTCS of the hierarchies with and without the LLM-based verification step on all datasets.}
\label{tab:LLMasJudge}
\end{minipage}
\end{figure*}
\textbf{Ablation study.}\label{sec:ablationStudy}
{\textit{(i) Effect of prompt constraints.}
{We first study how} incorporating different constraints, namely, the original instruction (Ori), and the variants ABS, FET, SLC and FULL (as illustrated in Figure~\ref{fig:P1Prompts}) affects \methodG’s performance (c.f. Section~\ref{sec:perTableHier}). }
As shown in Figures~\ref{fig:fig1} and~\ref{fig:fig2}, introducing any form of constraint generally improves performance across all PTCS and RI. In particular, applying FET or FULL yields substantial gains in top-level type inference compared to unconstrained prompts, with improvements of roughly 7\% on WDC and 5–16\% on GDS. Meanwhile, using SLC or ABS leads to more pronounced enhancements in the consistency metric PTCS of the generated hierarchies, achieving improvements of about 8–12\% on WDC and 5–10\% on GDS. 

\noindent{\textit{(ii) Effect of LLM-based verification.}
We evaluate the LLM-based verification step for pruning erroneous \textit{is-a} relationships by comparing \methodG with and without this step (c.f. Section~\ref{sec:mergePrune}). 
PTCS is reported because it measures the structural consistency between the inferred and ground-truth taxonomies. 
As shown in Table~\ref{tab:LLMasJudge}, verification improves PTCS on all datasets, with larger gains on WDC and OD Large (+0.153 and +0.135) and more moderate gains on GDS and OD Small (+0.053 and +0.078). 
These consistent improvements confirm that the verification step helps reduce erroneous \textit{is-a} relationships introduced during local path generation and merging.}

\subsection{Results on Conceptual Attributes}
\label{sec:eval-attributes}

The results for the conceptual attributes are presented in Table~\ref{tab:attriResults}. First, comparing \methodG to \methodE, we find that even our lowest performing \methodG variant achieves a comparable {RI} while delivering higher {Purity} (e.g., +0.18 on WDC) against \methodE's best performing setting. This suggests \methodG's methodological strength in forming semantically coherent clusters. Second, in a comparison against SI-LLM using the same LLM backbone (e.g., Qwen2.5-14B), \methodG consistently outperforms SI-LLM across all datasets, showing gains in RI (up to +0.062) and substantial improvements in {Purity} (up to 0.209). Overall, 
\methodG demonstrates its cross-domain robustness by consistently achieving high-accuracy results ({RI scores}~$>$~0.98 and {Purity}~$>$~0.9) across all tested datasets.


\begin{table}[tb]
\centering
\resizebox{\linewidth}{!}{
\begin{tabular}{l|ll|ll|ll}
\hline
\multirow{2}{*}{Method} & \multicolumn{2}{l|}{WDC}        & \multicolumn{2}{l|}{GDS}              & \multicolumn{2}{l}{OD Small}        \\ \cline{2-7} 
                        & RI             & Purity         & RI             & Purity               & RI                   & Purity               \\ \hline
\methodG (Llama-3.1-8B)    & 0.978          & 0.892          & 0.993          & 0.842                & 0.992                & 0.867                \\
\methodG  (Qwen2.5-14B)    & \underline{0.987} & \underline{0.927} & \underline{0.995}    & 0.894                & 0.993                & 0.906                \\
\methodG  (GPT-3.5)   & { 0.981}    & {0.923}    & \textbf{0.997} & {\underline{0.928}} & \textbf{0.995}       & \underline{{0.916}} \\
\methodG (GPT-4)           & \textbf{0.989}          & \textbf{0.931}          & 0.993          & \textbf{0.933}       & \underline{ 0.994}          & \textbf{0.923}       \\ \hline

\methodE (SBERT)                   & \textbf{0.933}          & \underline{0.686}          & \textbf{0.991}          & \underline{0.771}                & \textbf{0.994}       & \underline{0.787}          \\
\methodE (DeepJoin)                & \underline{0.898}          & 0.61           & 0.98           & 0.644                & 0.962                & 0.673                \\
\methodE (Unicorn)                 & 0.785          & \textbf{0.712}          & -              & -                    & 0.812                & 0.754                \\
\methodE (Starmie)                 & 0.873          & 0.497          & 0.976          & 0.641                & 0.944                & 0.633                \\
\methodE (SwAV)                    & \textbf{0.933}          & 0.664          & \underline{0.989}          & \textbf{0.803}                & \underline{0.983}          & \textbf{0.798}       \\ \hline
SI-LLM (Qwen2.5-14B)      & \textbf{0.941}          & \textbf{0.742}          & \textbf{0.98}           & \textbf{0.711}                & 0.931                & \underline{0.697}          \\
SI-LLM (GPT-3.5)          & \underline{0.916}          & 0.703          & 0.958          & 0.478                & {\textbf{0.966}} & 0.653                \\
SI-LLM (GPT-4)            & 0.908          & \underline{0.734}          & \underline{0.966}          & \underline{0.65}                 & \underline{0.964}          & \textbf{0.703}       \\ \hline
\end{tabular}
}
\caption{Results of the conceptual attributes. The best result of different settings for each method is bolded, and the second best is underlined.}
\label{tab:attriResults}
\end{table}

\methodG's superior performance can be attributed to its two-stage mechanism, which is underpinned by the {inferred column type hierarchy} (detailed in Section~\ref{sec:cthc}). This hierarchy allows \methodG to form more coherent initial column clusters by preventing coarse merges. For instance, in Figure~\ref{tab:musicRecording} and~\ref{tab:book}, it can correctly form a fine-grained \textit{musician} column cluster (grouping \textit{composer} and \textit{performer} columns) while properly excluding conceptually similar columns like \textit{author} from Figure~\ref{tab:book} about \textit{Book}, which \methodE and SI-LLM tend to merge into one broad \textit{person} cluster. The later attribute name inference and resolution step also improves the ability of \methodG to infer distinct semantic roles, which further separates \textit{musician} into \textit{composer} and \textit{performer} in the example, reduces inconsistencies, and enables the fine-grained, semantically consistent attribute grouping reflected in \methodG's high Purity scores.


Despite its robustness, \methodG remains challenged by {numeric and acronym-based columns} that lack informative column names or discriminative values. For instance, in \textit{FIFA World Cup} competition tables, columns such as \textit{Right of Center Midfielder} and \textit{Striker}, abbreviated as ``\textit{rcm}'' and ``\textit{st}'', contain homogeneous numeric scores and limited contextual cues. Consequently, \methodG fails to disambiguate them and incorrectly merges them into a single, generic \textit{Number} cluster. This illustrates the method's limitations when distinct semantic attributes lack sufficient contextual signals.

SI-LLM performs attribute inference by generating attribute names and resolving them through an LLM-based clustering method.  
While effective for high-level grouping, this approach tends to over-generalize and mix semantics: for example, \textit{game\_identifier}, \textit{game\_title}, and \textit{game\_description} are often collapsed into \textit{Game Info}, whereas distinct numeric attributes such as \textit{number\_of\_user\_ratings} and \textit{number\_of\_alternates} are incorrectly merged due to lexical similarity.  
These cases highlight the limitations of purely name-based reasoning when applied to large and semantically diverse attributes.


In contrast, \methodE is more effective on columns with rich descriptive text, but it frequently merges columns from different conceptual domains that share lexical or structural patterns.
For example, in \textit{Hotel} tables, the \textit{name} columns include values such as ``\textit{The Queen’s Gate Hotel}'' and ``\textit{Opera House Hotel}''; in \textit{Opera House} tables, the \textit{name} columns include ``\textit{Queen’s Theatre London}'' and ``\textit{The Metropolitan Opera}''.
This lexical overlap (e.g., ``\textit{Queen's}'', ``\textit{Opera}'') is the root of the confusion. Because the column embeddings used by \methodE are heavily influenced by this surface-level word overlap, their embeddings are similar. \methodE therefore incorrectly merges these conceptually distinct `name` columns, failing to distinguish their true conceptual roles (i.e., one is about Hotels, the other is about Theatre).

Among the embedding-based variants, EmSI (SwAV) achieves the highest Rand Index and Purity on GDS and OD Small.
This is because the attribute column clusters in these datasets exhibit a long-tailed distribution (i.e., many small, fine-grained column clusters). 
SwAV's fine-tuning process works by comparing 'column views' (augmented versions of columns) across training batches and pulling the semantically similar ones closer together.
This creates dense and well separated clusters. Consequently, SwAV excels at preserving small, fine-grained groups, preventing them from being incorrectly merged into larger, semantically mixed groups.
In contrast, EmSI (Unicorn) failed to scale to large datasets like GDS. Its quadratic pairwise computations required more GPU memory than was available.

\subsection{Results on Relationships and their Cardinalities}
\label{sec:eval-relationships}

\begin{table*}[htb]
\centering
\begin{minipage}[t]{0.6\textwidth}
\centering
\small
\resizebox{\linewidth}{!}{%
\begin{tabular}{c|ccc|ccc|lll}
\hline
\multicolumn{1}{l|}{\multirow{2}{*}{Method}} & \multicolumn{3}{c|}{{WDC}}     & \multicolumn{3}{c|}{{GDS}}     & \multicolumn{3}{c}{{OD Small}}                                                     \\ \cline{2-10} 
\multicolumn{1}{l|}{}                        & {P} & {R} & {F1} & {P} & {R} & {F1} & \multicolumn{1}{c}{{P}} & \multicolumn{1}{c}{{R}} & \multicolumn{1}{c}{{F1}} \\ \hline
\methodG (Llama-3.1-8B) & 0.696      & 0.655      & 0.675       & 0.716      & 0.684      & 0.700       & 0.792                          & 0.786                          & 0.785                           \\
\methodG (Qwen2.5-14B)                            & \textbf{0.768}      & \underline{0.7}        & \underline{0.733}       & 0.743      & \underline{0.743}      & {0.743}       & 0.875                          & \underline{0.929}                          & 0.900                           \\
\methodG (GPT-3.5)  & \underline{0.767}      & \textbf{0.733}      & \textbf{0.75}        & \textbf{0.843}      & 0.68       & \underline{0.753}       & \underline{0.938}     & \textbf{1.000}    & \textbf{0.967}  \\
\methodG (GPT-4)    & 0.725      & \textbf{0.733}      & 0.729       & \underline{0.804}      & \textbf{0.809}      & \textbf{0.807}       & \textbf{1.000}                          & \underline{0.929}                          & \underline{0.962}                           \\ \hline
\methodE (SBERT)    & \textbf{0.737}      & \underline{0.221}      & \underline{0.34}        & \textbf{0.667}    & 0.097  & \underline{0.169} & \textbf{0.500} & 0.143  & \textbf{0.222}    \\
\methodE (DeepJoin)   & \underline{0.684}      & \textbf{0.243}      & \textbf{0.359}       & \underline{0.624}      & \underline{0.103}      & \textbf{0.177}       & \underline{0.333}                          & 0.143                          & \underline{0.200}                           \\
\methodE (Unicorn)                                      & -          & -          & -           & -          & -          & -           & -                              & -                              & -                               \\
\methodE (Starmie)                                      & 0.25       & 0.115      & 0.158       & 0.121      & \textbf{0.184}      & 0.146       & 0.200                          & 0.143                          & 0.167                           \\
\methodE (SwAV)                                         & 0.152      & 0.204      & 0.174       & 0.143      & 0.05       & 0.074       & 0.125                          & 0.143                          & 0.133                           \\ \hline
\end{tabular}
}
\caption{Results of the relationships. The best result of different settings for each method is bolded, and the second best is underlined.}
\label{tab:relationshipResults}
\end{minipage}
\hfill
\begin{minipage}[t]{0.38\textwidth}
\centering
\small
\resizebox{\linewidth}{!}{
\begin{tabular}{l|l|l|l}
\hline
Method             & WDC   & GDS   & OD Small \\\hline
GeSI (Llama-3.1-8B) & 0.407 & 0.817 & 0.429            \\
GeSI (Qwen2.5-14B) & 0.870 & 0.712 & 1                \\
GeSI (GPT-3.5)     & 0.630 & 0.525 & 0.429            \\
GeSI (GPT-4)       & 0.804 & 0.847 & 0.857          \\  \hline   
\end{tabular}
}
\caption{Results of the relationship cardinalities by \methodG}
\label{tab:card}
\end{minipage}
\end{table*}

\textbf{Relationships.} Table~\ref{tab:relationshipResults} reports the results of the relationships.  
%
%
From the table, \methodG consistently outperforms \methodE on {F1-score}, even comparing \methodG's worst variant against \methodE's best. \methodE's variants exhibit a severe precision-recall imbalance. While they can achieve moderate {Precision} (e.g., 0.737 on WDC, 0.667 on GDS), this comes at the cost of low {Recall} (e.g., 0.243 on WDC, 0.184 on GDS). This imbalance causes its F1-score to collapse; \methodE's best F1-score remains low across all datasets, peaking at just 0.359 on WDC.
Conversely, even the worst-performing \methodG variant achieves a more balanced and higher performance. It maintains comparable or higher {Precision} while delivering considerably higher {Recall}, exceeding \methodE's best recall by +0.500 on GDS and +0.643 on OD Small.
This translates to a clear advantage in F1-score, where \methodG's worst variant outperforms \methodE's best by +0.316 on WDC, +0.523 on GDS, and +0.563 on OD Small.


While \methodE can effectively capture {instance-level evidence}, it primarily tests whether columns are similar.
Consequently, it struggles with conceptual-level relationships where overlapping column instances do not imply a true type-level relationship.  
For example, values like ``\textit{The Metropolitan Opera}'' and ``\textit{Hanguk Arts}'' in the columns of the attribute \textit{Event.organizer} may lead \methodE to falsely connect \textit{Event} with \textit{Opera House}. This example illustrates the sensitivity of \methodE to instance-level evidence, which may not always align with the desired conceptual-level relationship.



\methodG attempts to bridge this gap by jointly utilizing the attribute name and its sampled values to infer inter-type relationships, leading to marked improvements in Recall and F1.
However, it is sensitive when the sampled cell values for a single attribute represent multiple, distinct underlying entity types.
For instance, the attribute \textit{CreativeWork.Artist} often contains both individuals (e.g., ``\textit{John Lennon}``) and group artists (e.g., ``\textit{The Beatles}``, ``\textit{Queen}``, ``\textit{Pink Floyd}``).
If the sampled cell values for this attribute include some of these group artists (e.g., ``\textit{Queen}``, ``\textit{Pink Floyd}``) alongside individuals, the LLM may recognize that these specific group-artist values can be viewed as instances of \textit{Organization}.
The model then incorrectly generalizes from these specific values, inferring a relationship between the entire \textit{CreativeWork.Artist} attribute and \textit{Organization}, even though the attribute also contains many individuals.
These semantically reasonable but structurally inaccurate links partly explain the lower precision observed in relationship inference.

\textbf{Relationship cardinalities.} \label{sec:card}
We further evaluate \methodG on the task of inferring \textit{relationship cardinalities} after inter-type relationships are established.  
%
%
As shown in Table~\ref{tab:card}, \methodG achieves high agreement with the GT across all the datasets, yet performance varies across LLMs.  
Among them, {GPT-4} provides the most consistent results (from 0.80 to 0.85), while {Qwen2.5-14B} attains the highest accuracy on WDC (0.87) and perfect alignment on OpenData (1.00), but drops notably on GDS (0.71).  
These variations reveal distinct reasoning behaviors: GPT-4 effectively integrates statistical cues with contextual semantics, whereas Qwen relies more on statistics and evidence in the provided table samples.  
Overall, the results suggest that incorporating sampled table instances and column-level statistics improves the model’s capacity to infer relationship cardinalities.

Nevertheless, these gains do not fully mitigate model sensitivity to incomplete or asymmetric samples.  
For example, in Table~\ref{tab:book}, the author \textit{Alexandre Dumas fils} is linked to books such as \textit{Diane de Lys} and \textit{La Traviata}.  
When only one of these works appears in the sample, the model may incorrectly infer a \textit{1:n} (\textit{Book}→\textit{Author}) relationship instead of the correct \textit{m:n}, reflecting the model's tendency to confuse the observed cardinality in the sample (which appears to be 1:n) with the true cardinality of the conceptual schema (which is m:n).


\subsection{Efficiency and {Cost Analysis}}
\begin{figure*}[tb]
  \centering
  \begin{minipage}[b]{0.31\textwidth} 
    \centering
    \includegraphics[width=\textwidth]{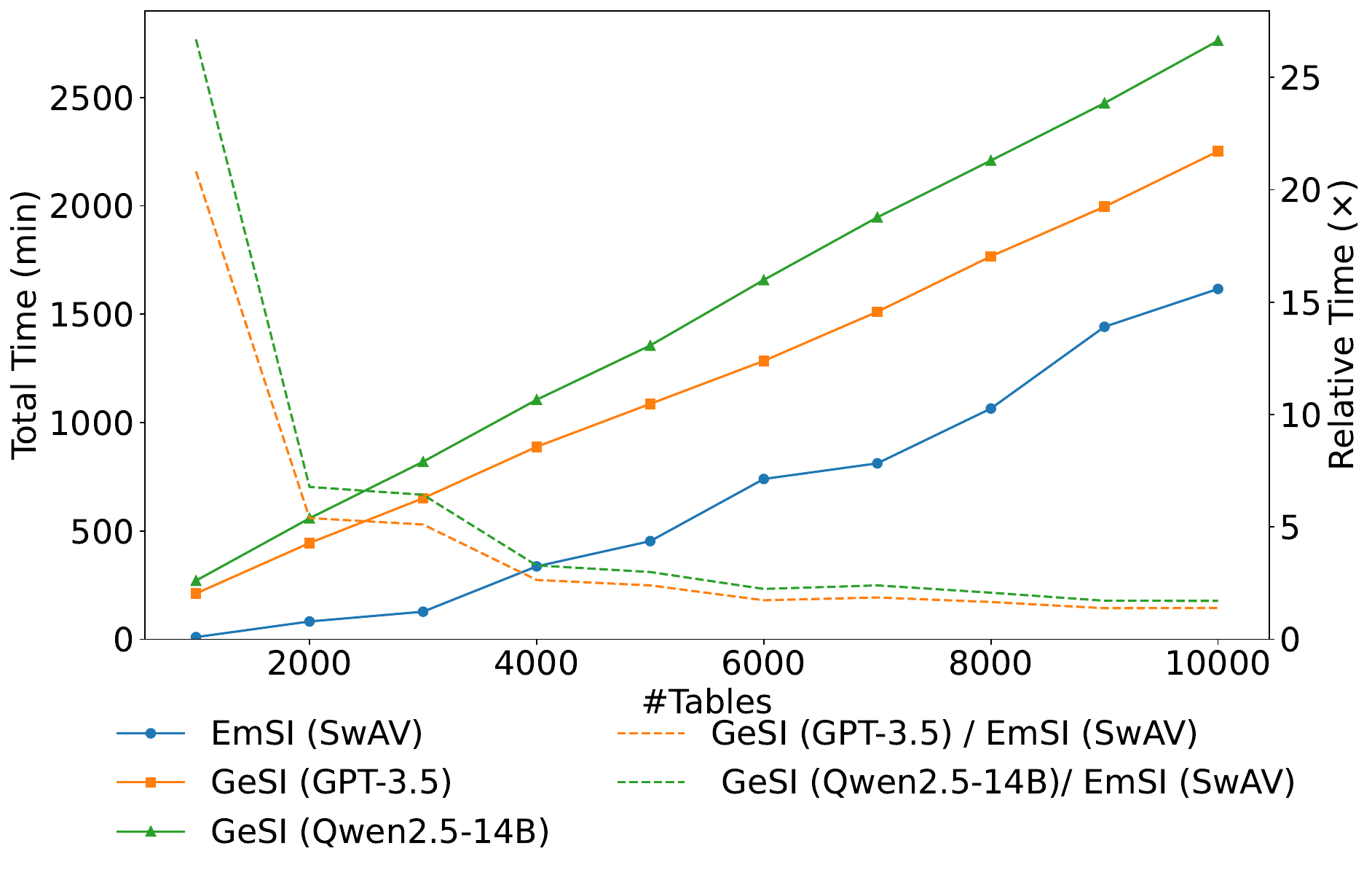}
    \caption{Overall runtime of \methodG(Qwen2.5-14B/ GPT-3.5) and \methodE (SwAV) on {OD Large}}
    \label{fig:overalleffi}
 \end{minipage}
 \begin{minipage}[b]{0.315\textwidth}
  \centering
    \includegraphics[width=\textwidth]{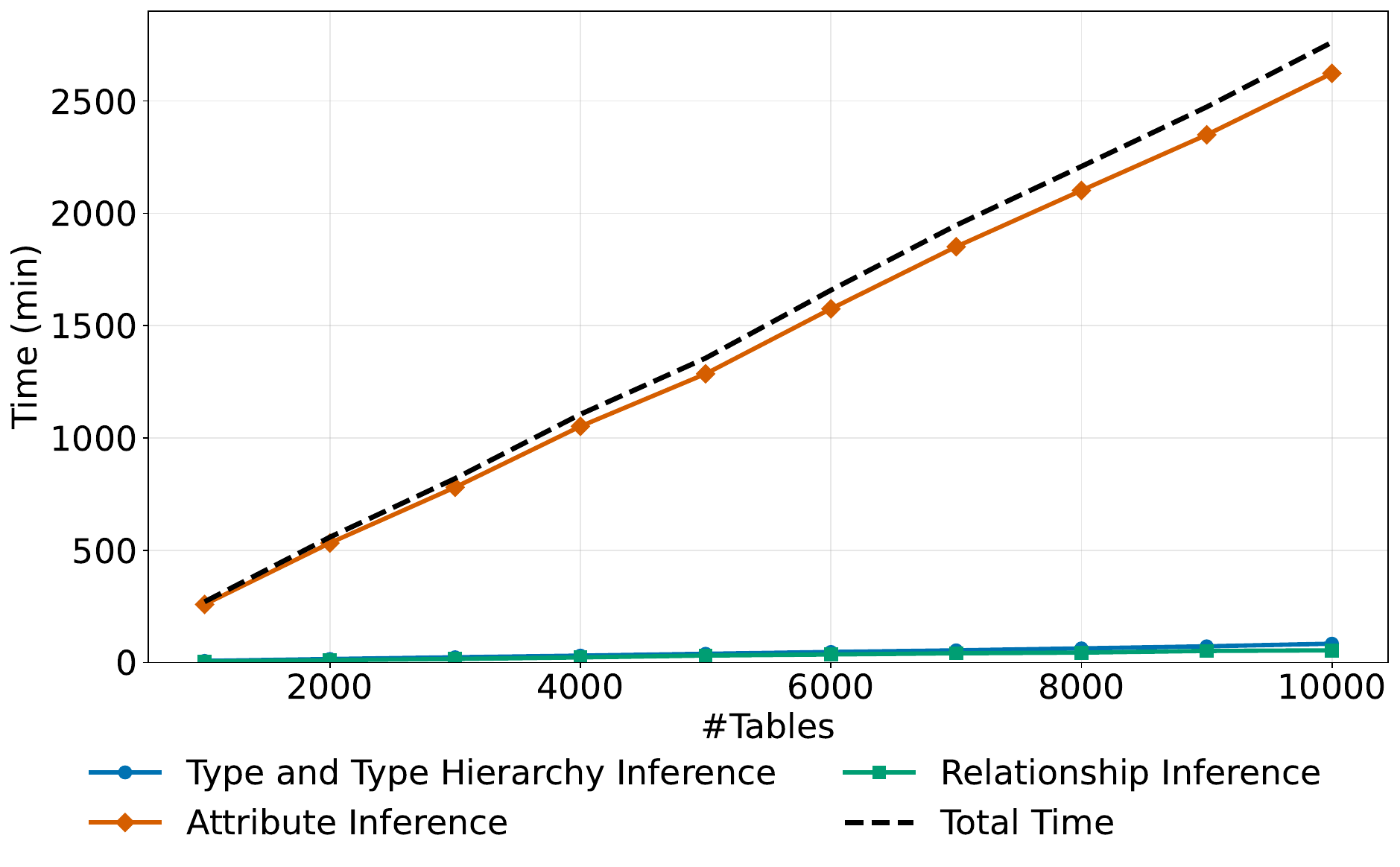}
    \caption{Overall runtime and step-wise runtime for \methodG (Qwen2.5-14B) on OD Large}
    \label{fig:gesi_breakdown}
    \end{minipage}
 \begin{minipage}[b]{0.33\textwidth}
  \centering
    \includegraphics[width=\textwidth]{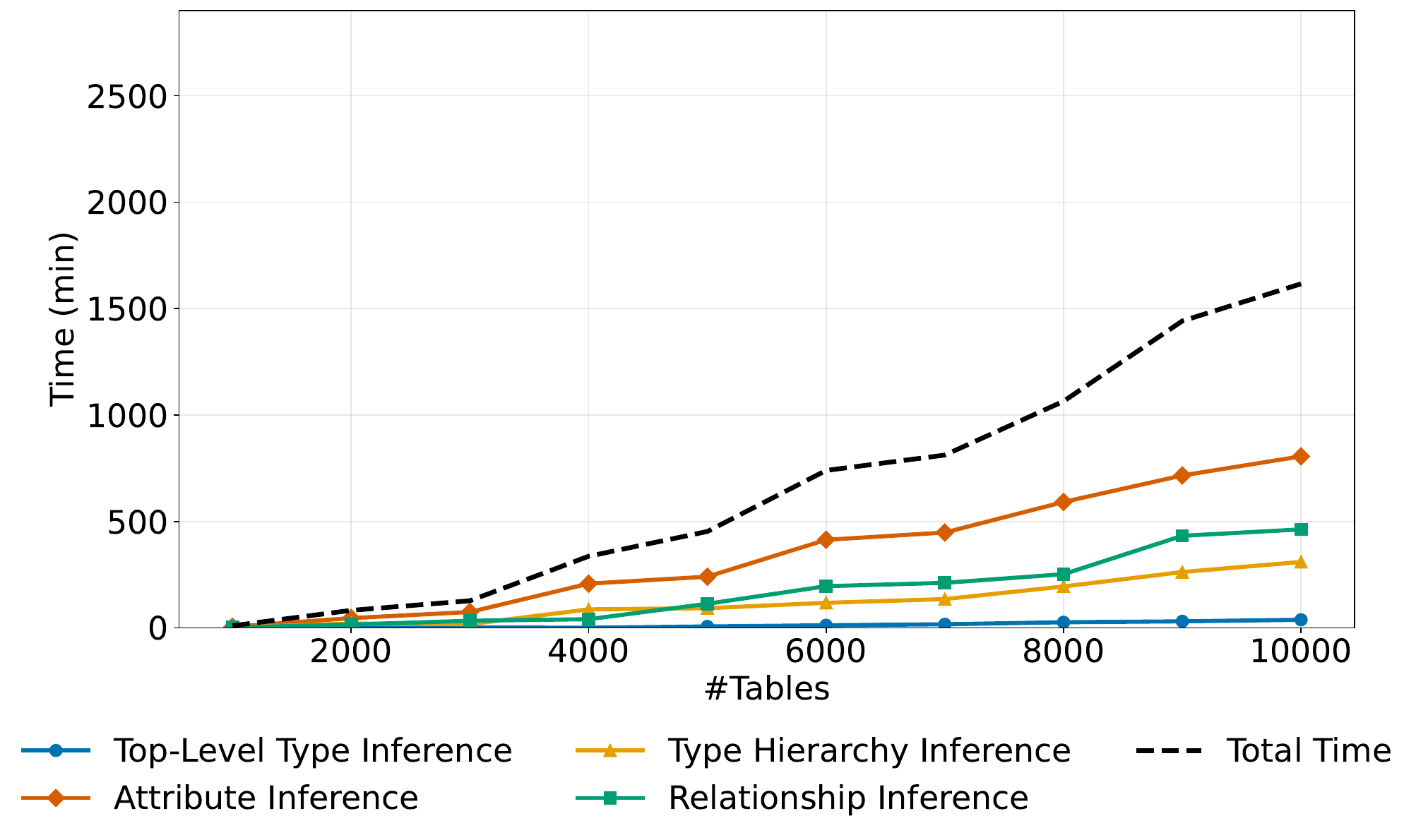}
    \caption{Overall runtime and step-wise runtime for \methodE (SwAV) on OD Large}
    \label{fig:emsi_breakdown}
    \end{minipage}
\end{figure*}

\textbf{Overall runtime.} Figure~\ref{fig:overalleffi} presents the overall runtime comparison between \methodE (using SwAV) and \methodG (using Qwen2.5-14B and GPT-3.5) on the OD Large dataset, executed on an NVIDIA A100 GPU (80 GB).  
Runtimes are measured across varying scales, with the number of tables (resp. columns) increasing from 1K to 10K (resp. from 16K to 355K).  
For \methodG, we report both open-source (Qwen2.5-14B) and closed-source (GPT-3.5) implementations.  
The total runtime (solid lines) and relative runtime ratios (dashed lines, showing \methodG's performance relative to \methodE's) are plotted as a function of the number of processed tables.

At small to medium scales (1K–3K tables), \methodE shows a pronounced efficiency advantage, requiring less than 5\% of \methodG’s total runtime. As the dataset size increases, the relative gap decreases steadily; at 10K tables, \methodG (GPT-3.5) is only about 1.4× slower than \methodE. Both methods exhibit approximately linear growth w.r.t the number of tables, indicating near-$O(n)$ scalability.  
\methodG displays smoother scaling behavior owing to the uniform batching of LLM calls, whereas \methodE shows a slightly steeper slope, suggesting that its fixed-cost advantage diminishes at larger scales.  
Between the two generative variants of \methodG, GPT-3.5 consistently outperforms Qwen2.5-14B by roughly 20–25\%, reflecting differences in underlying inference throughput.

\noindent\textbf{Step-wise runtime.} Figures~\ref{fig:gesi_breakdown} and~\ref{fig:emsi_breakdown} illustrate the runtime of each step in \methodG (Qwen2.5-14B) and \methodE (SwAV) w.r.t. the number of tables tested over the OD Large dataset.
%
As shown in Figure~\ref{fig:gesi_breakdown}, for \methodG, the {type and hierarchy inference} and {conceptual attribute inference} phases dominate the total runtime, with the latter accounting for over 90\% of the overall runtime.  
In contrast, {type and hierarchy inference} contributes only a small fraction of the total time, as it involves a single generative inference step per batch of tables.

The overall runtime of \methodE is dominated by three phases: {attribute inference}, {type hierarchy inference}, and {relationship inference}. In contrast, {top-level type inference} is lightweight, as its empirical cost grows slowly and contributes minimally to the total runtime. 
Attribute inference constitutes the largest portion of the runtime. This is because its clustering process operates over columns ($m$), which scale to hundreds of thousands, leading to a quadratic complexity ($O(m^2)$).
The runtime of type hierarchy inference also grows quadratically ($O(n^2)$), due to pairwise similarity computations between tables ($n$). However, attribute inference remains the primary bottleneck, as its complexity scales with the number of columns ($m$), which in our datasets can be orders of magnitude larger than the number of tables ($n$). In comparison, the cost of relationship inference increases moderately, driven by iterations over attribute pairs both within and across types.
\begin{table}[tb]
\centering
\small
\setlength{\tabcolsep}{4pt}
\renewcommand{\arraystretch}{0.95}
\resizebox{0.85\linewidth}{!}{%
\begin{tabular}{lrrrrr}
\toprule
\multirow{2}{*}{Dataset} 
& \multirow{2}{*}{Calls} 
& \multicolumn{4}{c}{Cost (USD)} \\
\cmidrule(lr){3-6}
& & Type & Attr. & Rel./Card. & Total \\
\midrule
WDC      & 5450  & 1.742 & 1.290 & 0.050 & 3.082 \\
GDS      & 17006 & 2.100 & 1.550 & 0.120 & 3.770 \\
OD Small & 4420  & 0.645 & 0.540 & 0.043 & 1.228 \\
\bottomrule
\end{tabular}}
\caption{Estimated financial cost of \methodG.}
\label{tab:gesi_cost_analysis}
\end{table}

\noindent\textbf{Cost analysis.}
{We estimate the financial cost of the GPT-based \methodG which calls OpenAI API. 
For each dataset, we record the number of LLM calls and the input/output token usage in the main inference stages, and estimate the USD cost using GPT-3.5 as a representative low-cost closed-source model. 
Table~\ref{tab:gesi_cost_analysis} shows that the total cost is \$3.082 for WDC, \$3.770 for GDS, and \$1.228 for OD Small, with most of the cost coming from type-path and attribute inference.}



 \section{Case Study}
\label{sec:case} 

\begin{figure*}[htbp]
  \centering
  \begin{subfigure}[t]{0.48\textwidth}
    \centering
      \resizebox{\textwidth}{!}{ 
    \includegraphics[width=\linewidth]{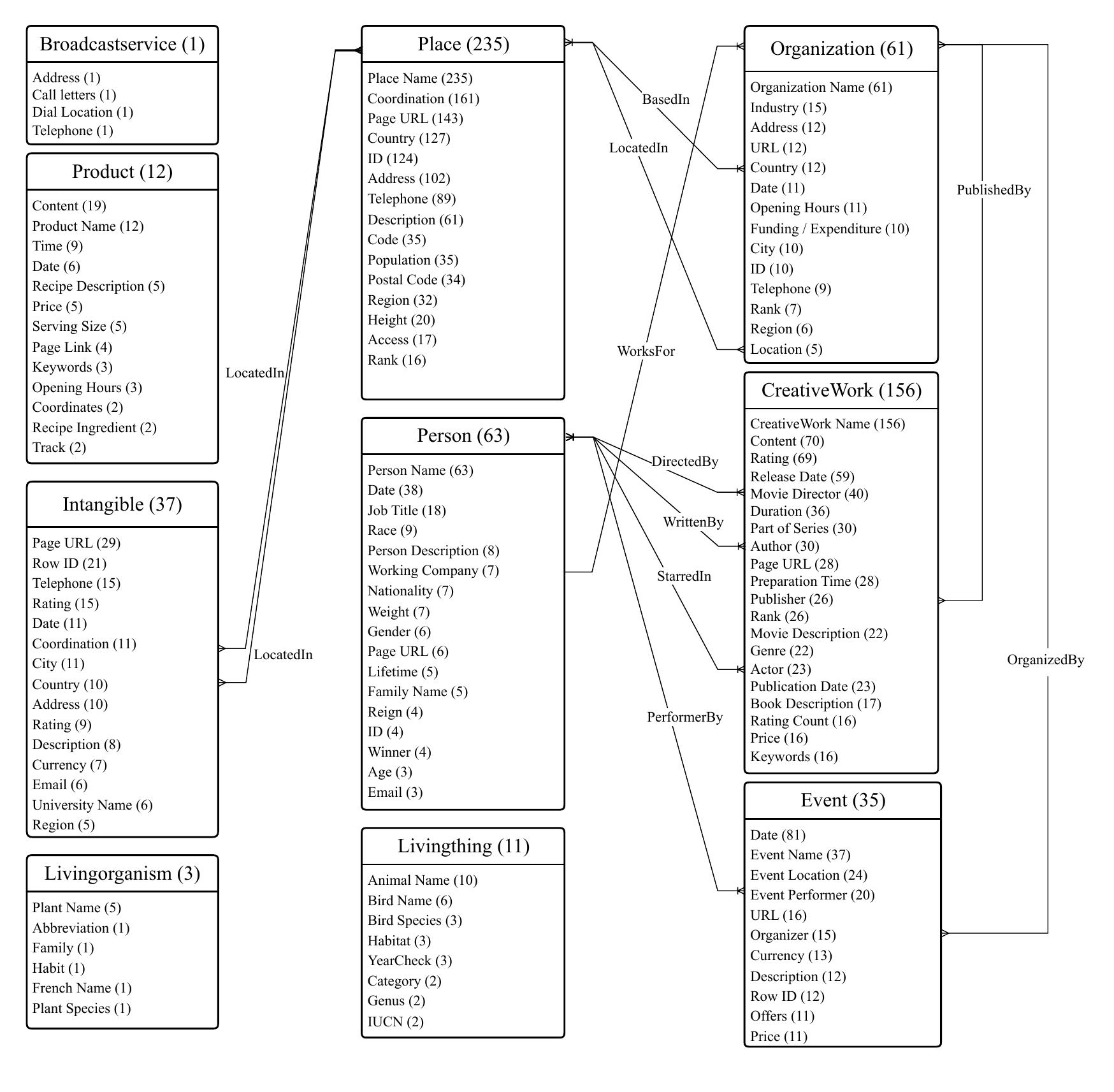}}
    \caption{Inferred top-level types by \methodG}
    \label{fig:gesi-top}
  \end{subfigure}
  \begin{subfigure}[t]{0.48\textwidth}
    \centering
     \resizebox{0.9\textwidth}{!}{ 
    \includegraphics[width=\linewidth]{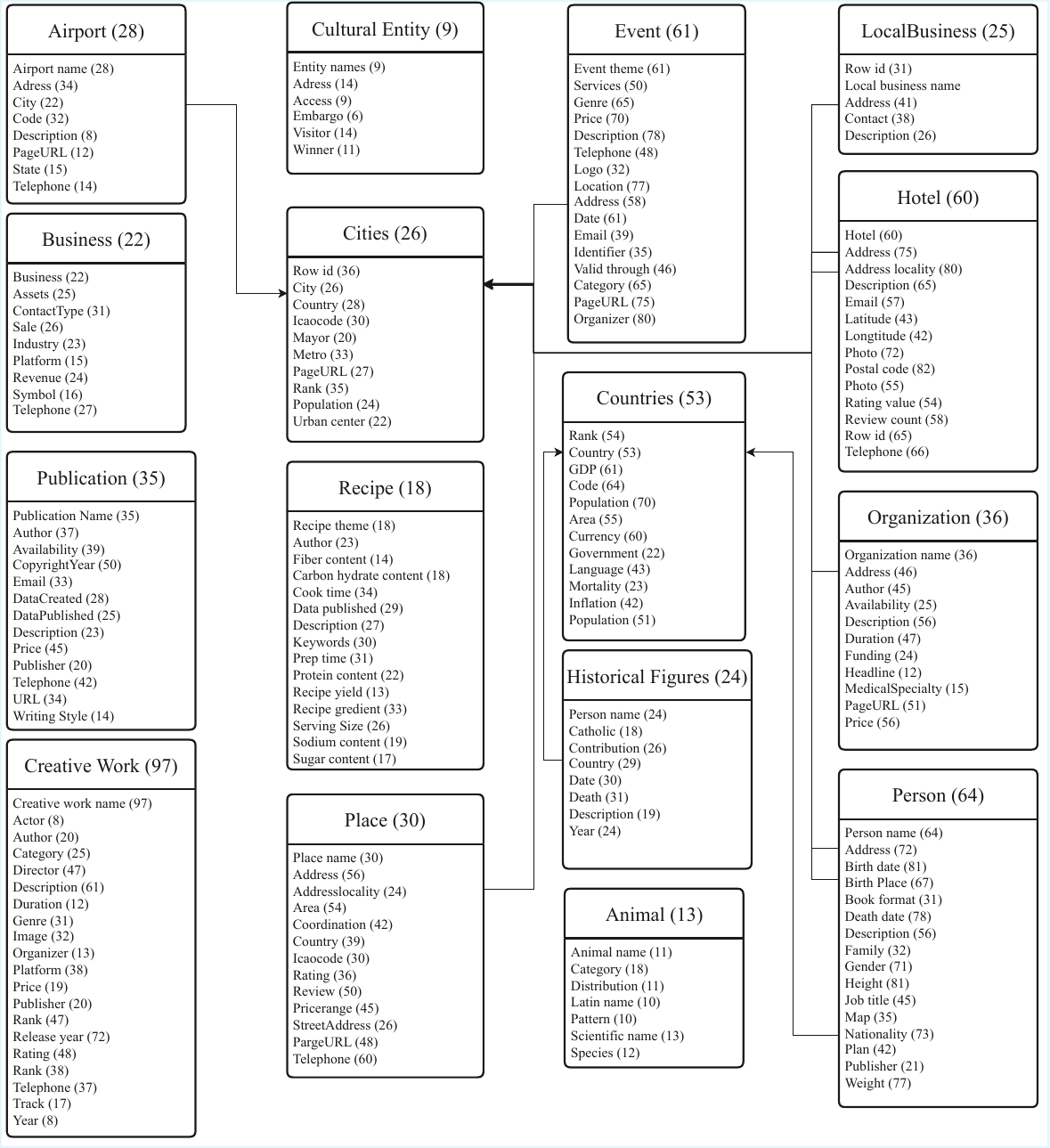} 
    }
  \caption{Inferred top-level types by \methodE}
  \label{fig:emsi-top}
  \end{subfigure}
  \begin{subfigure}[t]{0.8\textwidth}
    \centering
      \resizebox{\textwidth}{!}{ 
    \includegraphics[width=\linewidth]{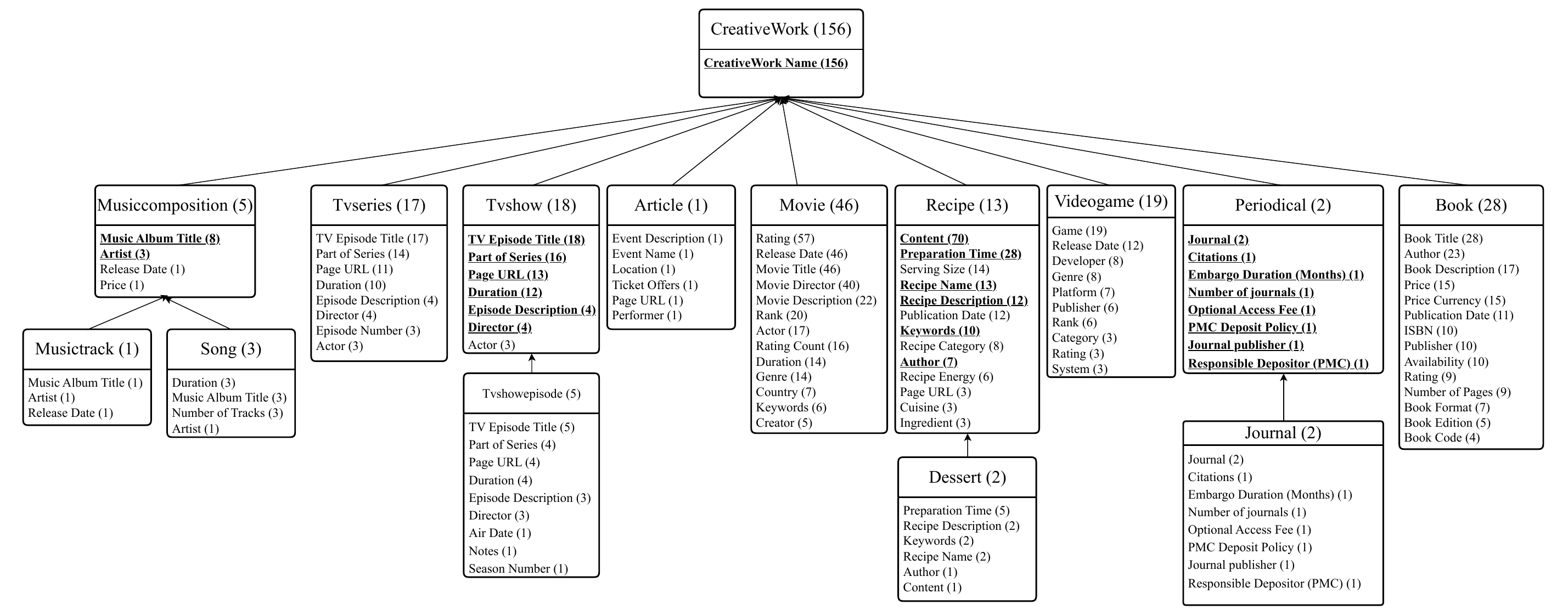}
    }
    \caption{Sub-hierarchy under CreativeWork by \methodG}
    \label{fig:gesi-sub}
  \end{subfigure}

  \begin{subfigure}[t]{0.8\textwidth}
    \centering
    \resizebox{\textwidth}{!}{ 
    \includegraphics[width=\linewidth]{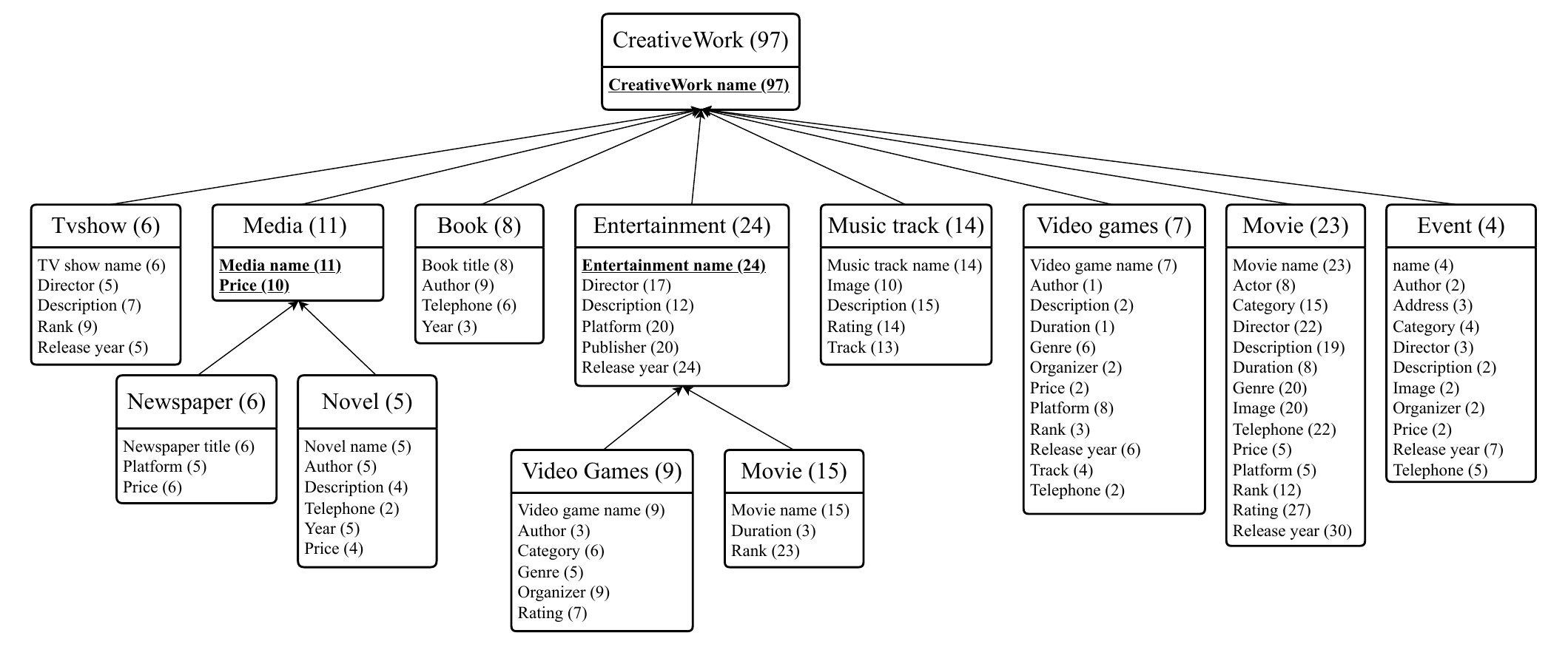}
    }
    \caption{Sub-hierarchy under CreativeWork by \methodE}
    \label{fig:emsi-sub}
  \end{subfigure}
  \caption{Comparison between \methodG and \methodE in top-level types and sub-hierarchy inference.}
  \label{fig:ETE}
\end{figure*}

To further evaluate the behavior of our schema inference methods \methodE and \methodG under real-world data, we present a case study from the WDC {dataset}, which exhibits a clear long-tailed distribution of types and their associated tables.
This setting provides an opportunity to assess how effectively the two methods infer the overall schema structure while distinguishing fine-grained types that occur infrequently.
As \methodE lacks a native naming component, we leverage {GPT-3.5} for post-hoc naming to facilitate visualization.

We visualize and compare the inferred schema components from both methods, focusing on:
(1) the {top-level types} together with their attributes and relationships (Figures~\ref{fig:gesi-top} and~\ref{fig:emsi-top}); and
(2) the {sub-hierarchy} under one inferred top-level type, \textit{CreativeWork}, along with its attributes (Figures~\ref{fig:gesi-sub} and~\ref{fig:emsi-sub}).
We select the \textit{CreativeWork} sub-hierarchy because it is identified as a top-level type by both methods and, in the GT schema, its sub-hierarchy exhibits a pronounced long-tail distribution.
For reference, the GT schema covers seven top-level types: \textit{Intangible} (8), \textit{Place} (188), \textit{Person} (65), \textit{Organization} (123), \textit{Organism} (14), \textit{CreativeWork} (166), and \textit{Event} (37), where the numbers represent the number of tables associated with each type.


\noindent\textbf{Top-level types.} 
Figures~\ref{fig:gesi-top} and~\ref{fig:emsi-top} show that the inferred schema of \methodG aligns closely with the GT schema, particularly for \textit{Event} (35 vs.\ 37), \textit{Person} (63 vs.\ 65), and \textit{CreativeWork} (156 vs.\ 166), demonstrating robustness under long-tailed sparsity. Overall, \methodG identifies 10 top-level types across 602 tables, capturing most top-level types in the corpus.
The enlarged \textit{Intangible} cluster results from the contextual merging of semantically mixed types such as \textit{Event}, while a similar overlap appears between \textit{Place} and \textit{Organization}, where location-based organizations, such as \textit{Hotel} and \textit{LocalBusiness} are grouped under \textit{Place}.
Occasionally, the model produces near-synonymous types (e.g., \textit{LivingThing} and \textit{LivingOrganism}), reflecting minor lexical variation rather than conceptual divergence.

The relationships inferred by \methodG primarily interconnect the \textit{CreativeWork}, \textit{Person}, \textit{Organization}, and \textit{Place} types. Illustrative examples include the \textit{DirectedBy} relation between \textit{CreativeWork} and \textit{Person} (inferred as m:n) and the \textit{WrittenBy} relation (also between \textit{CreativeWork} and \textit{Person}) (inferred as n:1). The model also identified other complex m:n relationships, such as that connecting \textit{Event} and \textit{Person}.

These relationships generally correspond to natural interactions among entities: people creating works, organizations operating in locations, and people participating in events. However, the inferred cardinality exhibits limitations. For instance, \methodG infers an n:1 cardinality for \textit{WrittenBy,} \textit{CreativeWork} and \textit{Person}, which over-simplifies the real m:n cardinality (as works can have multiple authors). This specific error likely results from the LLM's tendency to confuse the observed cardinality in the given table sample (which appears n:1 due to data sparsity) with the true conceptual cardinality (which is m:n), a limitation we analyzed in Section ~\ref{sec:card}. Overall, while not perfectly accurate in all cardinality inference, \methodG reconstructs a largely coherent and semantically rich schema that reflects how entities are interconnected across entity types.

In contrast, \methodE produces a schema with more top-level types (16 inferred from the 602 datasets). On one hand, this demonstrates sensitivity to distinct conceptual domains: types such as \textit{Hotel}, \textit{LocalBusiness}, \textit{Recipe}, and \textit{Publication} exhibit coherent table sets, showing that embedding-based clustering can effectively separate tables with distinct semantics. However, the inferred types vary widely in conceptual scope. Some (e.g., \textit{Event}, \textit{CreativeWork}) remain broad and multi-themed, while others (e.g., \textit{Animal}, \textit{City}) represent narrow types that would, in a hierarchy, fall under broader types such as \textit{Place} or \textit{Organism}.  
This uneven granularity makes the hierarchy in the schema unbalanced, as some branches are far more abstract than others.

The relationships identified by \methodE are primarily location-oriented, reflecting the limited availability of explicit relational evidence in the tables.  
Attributes participating in these relations share high mutual information in their cell values, suggesting that \methodE captures instance-based rather than conceptual relationships.  
Moreover, as described in Section~\ref{sec:infer-hierarchies}, its hierarchy construction connects tables through shared conceptual attributes instead of explicit parent-child semantics, implying that the model organizes related tables by instance-level semantic similarity rather than ontological structure.

\noindent\textbf{Sub-hierarchy under CreativeWork.}  
As shown in Figure~\ref{fig:gesi-sub}, the sub-hierarchy inferred by \methodG under \textit{CreativeWork} is structurally consistent, containing no unsuitable is-a relationships between types.  
It successfully identifies several long-tail types. For instance, \textit{Journal} and \textit{TVshow}, which in the GT are supported by only two and one tables, respectively,
highlighting its effectiveness for inferring types even with minimal supporting data.
However, several sibling types display overlapping or uneven naming granularity (e.g., \textit{Musictrack} and \textit{Song}), indicating that \methodG sometimes treats near-synonyms as separate sibling types. 

The hierarchy inferred by \methodE under \textit{CreativeWork} is compact and conceptually coherent. 
As illustrated in Figure~\ref{fig:emsi-sub}, it captures major subtypes such as \textit{Movie}, \textit{Book}, and \textit{MusicTrack}, maintaining correct parent-child relationships without reversals.  
An unexpected subtype, \textit{Event}, appears under \textit{CreativeWork}; this arises because several event tables describe concert or performance data (e.g., \textit{organizer}, \textit{director}, \textit{release year}), whose attribute patterns resemble media tables, so \methodE incorrectly places them under \textit{CreativeWork}. 
Compared with \methodG, \methodE yields a shallower hierarchy that emphasizes frequent types while omitting rare ones such as \textit{Journal} and \textit{Recipe}.  

\section{Conclusion}
\label{sec:conclu} 


A conceptual schema can play an important role for managing and exploring large and heterogeneous tabular datasets but its inference is challenging. In this work, we infer a conceptual schema consisting of types, type hierarchies, conceptual attributes and relationships between types, under a minimal metadata assumption, by developing two LLM-based approaches: \methodG, a generative method using a decoder-based LLM and task-specific prompts, and \methodE, a semantic similarity-based method using table embeddings from an LLM encoder. 
Extensive experiments on four ground truth annotated benchmarks constructed from web and open data validate that \methodE and \methodG, as well as each of their steps, can achieve promising results.
Beyond the empirical results, this study highlights the potential of language models to connect table-level data with conceptual abstractions, providing a concise and interpretable representation of the data in complex repositories.



\section*{Statements and Declarations}
\textbf{Competing Interests} The authors have no competing interests to declare that are relevant to the content of this article.\\
\textbf{Funding} Zhenyu Wu was supported by A Dean's Doctoral Scholarship from The University of Manchester. \\

\bibliographystyle{plain}

\bibliography{vada}



%
%

\end{document}